\documentclass[twocolumn]{aastex631}
\usepackage{lineno}
\usepackage{subfigure} 
\usepackage[para]{threeparttable}

\received{July 17, 2022}
\revised{September 7, 2022}
\accepted{November 4, 2022}

\shorttitle{Stellar Loci VI: An Updated Catalog of the Best and Brightest Metal-poor Stars}
\shortauthors{Xu et al.}

\begin{document}

\title{Stellar Loci VI: An Updated Catalog of the Best and Brightest Metal-poor Stars}

\author[0000-0003-3535-504X]{Shuai Xu}
\affiliation{Institute for Frontiers in Astronomy and Astrophysics, Beijing Normal University,  Beijing 102206, China}
\affiliation{Department of Astronomy, Beijing Normal University No.19, Xinjiekouwai St, Haidian District, Beijing, 100875, P.R.China}

\author[0000-0003-2471-2363]{Haibo Yuan}
\affiliation{Institute for Frontiers in Astronomy and Astrophysics, Beijing Normal University,  Beijing 102206, China}
\affiliation{Department of Astronomy, Beijing Normal University 
No.19, Xinjiekouwai St,
Haidian District, Beijing, 100875, P.R.China}

\author[0000-0003-1863-1268]{Ruoyi Zhang}
\affiliation{Institute for Frontiers in Astronomy and Astrophysics, Beijing Normal University,  Beijing 102206, China}
\affiliation{Department of Astronomy, Beijing Normal University No.19, Xinjiekouwai St, Haidian District, Beijing, 100875, P.R.China}

\author[0000-0002-0389-9264]{Haining Li}
\affiliation{Key Lab of Optical Astronomy, National Astronomical
  Observatories, Chinese Academy of Sciences (CAS) \\
A20 Datun Road, Chaoyang, Beijing 100101, China}

\author[0000-0003-4573-6233]{Timothy C. Beers}
\affiliation{Department of Physics and Astronomy and JINA Center for the Evolution of the Elements
(JINA-CEE), University of Notre Dame, Notre Dame, IN 46556, USA}

\author[0000-0003-3250-2876]{Yang Huang}
\affil{South-Western Institute for Astronomy Research, Yunnan University, Kunming 650500, People’s Republic of China}

\correspondingauthor{Haibo Yuan}
\email{yuanhb@bnu.edu.cn}

\begin{abstract}
We employ Gaia, 2MASS, and ALLWISE photometry, as well as astrometric data from Gaia, 
to search for relatively bright very metal-poor ([Fe/H] $< -2.0$; VMP) giant star candidates using three different criteria: 1) the derived Gaia photometric metallicities from Xu et al. (2022), 2) the lack of stellar molecular absorption near 4.6 microns, and 3) their high tangential velocities. With different combinations of these criteria, we have identified six samples of candidates with $G <$ 15: the Gold sample (24,304
candidates), the Silver GW sample (40,157 candidates), 
the Silver GK sample (120,452 candidates), the Bronze G 
sample (291,690 candidates), the Bronze WK sample (68,526 
candidates), and the Low $b$ sample (4,645 candidates). 
The Low $b$ sample applies to sources with low Galactic lattitude, $|b| < 10^\circ$, while
the others are for sources with $|b| > 10^\circ$. 
By cross-matching with results derived from medium-resolution ($R \sim$ 1800) from LAMOST DR8, we establish that the success rate for identifying VMP stars is 60.1$\%$ for the Gold sample, 39.2$\%$ for the Silver GW sample, 41.3$\%$ for the Silver GK sample, 15.4$\%$ for the Bronze G sample, 31.7$\%$ for the Bronze WK sample, and 16.6$\%$ for the Low $b$ sample, respectively. 
An additional strict cut on the quality parameter $RUWE < 1.1$ can further increase the success rate of the Silver GW, Silver GK,
and Bronze G samples to 46.9$\%$, 51.6$\%$, and 29.3$\%$, respectively. Our samples provide valuable 
targets for high-resolution follow-up spectroscopic observations, and are made publicly available. 

\end{abstract}

\keywords{Population II stars; Milky Way stellar halo; Stellar abundances}

\section{Introduction} \label{sec:intro} 

Very metal-poor (VMP) stars are defined as having [Fe/H] $< -2.0$ (\citealt{beers2005}). They are crucial ``fossil probes" of the first nucleosynthesis events in the early universe, and thus play an unique role in studies of Population\,III stars. The VMP stars in the Galactic halo also provide important constraints on the formation and early evolution of the Milky Way. The extremely metal-poor (EMP; [Fe/H] $< -3.0$) and ultra metal-poor (UMP; [Fe/H] $< - $4.0) stars are thought to form several hundred million years after the Big Bang, and in addition provide probes of lithium production from Big Bang nucleosynthesis \citep{iwamoto2005} and the production of carbon by high-mass early generation stars \citep{heger2010,meynet2010,chiappini2013}.

Medium-resolution spectroscopy ($R \sim 1200-2000$) is often utilized to search for metal-poor stars, making use of the Ca ${\rm \uppercase\expandafter{\romannumeral2}}$ H and K absorption lines (\citealt{beers1985,beers1992,Christlieb2008}), or the 
Ca ${\rm \uppercase\expandafter{\romannumeral2}}$ 
triplet lines (\citealt{Fulbright2010}).
Large-scale spectroscopic surveys, such as the Sloan Extension for Galactic Understanding and Exploration (SEGUE, \citealt{yanny2009}; SEGUE-2, \citealt{Rockosi2022}) and the Large sky Area Multi-Object fiber Spectroscopic Telescope (LAMOST; \citealt{cui2012,deng2012,Zhao2012,liu2014}), provide the opportunity 
to identify VMP stars from among millions of medium- to intermediate-resolution ($R \sim 5000$) spectra
(e.g., \citealt{li2015a,li2015b,li2015c,li2018,Aguado2017,mardini2019,li2022,lamostdr8}).

VMP candidates can also be effectively identified using photometric data,
particularly those that include narrow- and intermediate-band filters designed for stellar-parameter determinations, such as the SkyMapper survey (\citealt{keller2007,keller2014,huang2022}), the Pristine survey (\citealt{Starkenburg2017,Aguado2019}), the Javalambre Photometric Local Universe Survey (J-PLUS; \citealt{Cenarro2019,Whitten2019,Galarza2022,yang2022}) and the Southern Photomeric Local Universe Survey (S-PLUS; \citealt{Mendes2019},\citealt{Whitten2021},\citealt{Placco2022}). Stellar metallicities can also be precisely determined down to [Fe/H] $\sim -2.5$ with broadband filters, if the data quality is sufficiently high (e.g., \citealt{yuan2015,an2020,zhang2021,xu2022}).

For a VMP candidate, particularly those selected exclusively by photometry, follow-up high-resolution spectroscopy is often used to confirm whether or not it is a true VMP star, to determine if it is chemically peculiar in some respect (e.g., if it is carbon-enhanced or exhibits enhanced neutron-capture elements), or in order to carry out more in-depth studies of its full elemental-abundance distribution. According to \cite{schlaufman2014}, it takes about four hours for a 6.5m-aperture telescope to obtain a R $\sim$ 25,000 spectrum with S/N $\sim$ 100 pixel$^{-1}$ at 400 nm for a VMP candidate with V $\approx$ 16. Bright VMP candidates have the huge advantage that follow-up high-resolution spectroscopy can be obtained in substantially less time or even with smaller-aperture telescopes. 
By making use of the fact that VMP stars generally lack strong molecular absorption near 4.6 microns, \citet{schlaufman2014} identify 11,916 bright ($V < 14$) metal-poor star candidates from the public, all-sky APASS optical, 2MASS near-infrared, and WISE mid-infrared photometry. Follow-up high-resolution spectroscopy shows about 20$-$36\% of their candidates have [Fe/H] $< -2.0$ (\citealt{schlaufman2014,Casey2015}). 

\citet{Limberg2021} have demonstrated that the use of kinematic parameters, such as the radial velocities and tangential velocities (based on proper motions and distance estimates), or their combination,  can dramatically improve the success rates for identification of likely VMP stars in several previous surveys.  Below we use a similar scheme as one of our search criteria.

Taking advantage of the unprecedented photometric quality of Gaia Early Data Release 3 (EDR3; \citealt{GaiaEDR32021}), we have previously obtained reliable photometric metallicty estimates for a magnitude-limited sample of 27 million stars with $G < 16$ (\citealt{xu2022}), including nearly 7 million giant stars. In this work, we combine Gaia photometric metallicities,  kinematics, and ALLWISE colors to identify large samples of relatively bright ($G < 15$) VMP giant stars. 

This paper is organized as follows. The data is introduced in Section \ref{sec:data}. We describe our methodologies in Section \ref{sec:method}. The candidate samples we identify are presented in Section \ref{sec:result}, along with their estimated success rates. Section \ref{sec:con} presents a brief summary.

\section{Data} \label{sec:data}

The data we employ are drawn from the 2MASS All-Sky Point Source Catalog (2MASS; \citealt{tmass2006}), the ALLWISE Source Catalog (ALLWISE; \citealt{wise2011,neowise2011}), and Gaia
EDR3 (\citealt{GaiaEDR32021}).

2MASS collected data covering 99.998\% of the celestial sphere in the near-infrared $J$ (1.25 $\mu$m), $H$ (1.65 $\mu$m), and $K_S$ (2.16 $\mu$m) bands, and obtained data for 471 million stellar objects. 
The ALLWISE program builds upon the work of the successful Wide-field Infrared Survey Explorer mission (WISE; \citealt{wise2011}), by combining data from the WISE cryogenic and NEOWISE \citealt{neowise2011}). The ALLWISE Source Catalog has provided photometry in four infrared bands: $W1$ (3.4 $\mu$m), $W2$ (4.6 $\mu$m), $W3$ (12 $\mu$m), and $W4$ (22 $\mu$m) for over 747 million objects. 

Gaia EDR3 has provided the best available photometric data to date, obtaining colors with unprecedented mmag precision, as well as parallaxes of unprecedented $\mu$mas precision for more than one billion stars over the entire sky. Using the color correction of \cite{Niu2021EDR3}, the $G$-magnitude correction of \cite{Yang2021}, the parallax correction of \cite{lindegren2021}, and careful reddening corrections using empirical color- and reddening-dependent coefficients, \cite{xu2022} have provided photometric metallicities for a magnitude-limited sample of 27 million FGK stars down to [Fe/H] $\sim -2.5$ with $10 < G \leq 16$, $|b| > 10^\circ$, and $E(B-V) \leq 0.5$\,mag, based on the empirical metallicity-dependent stellar locus determined from LAMOST DR7 (\citealt{LAMOST}). The metallicity catalog of \cite{xu2022} contains about 7 million giants.
Their typical metallicity errors are 0.2 -- 0.3 dex for sources with $G < 15$.  We compare the [Fe/H] from \cite{xu2022} with four other catalogs in Appendix \ref{sec:appendix}, and find good consistency in all cases. Note that the General Stellar Parametrizer from Spectroscopy of Gaia (Gaia GSP-Spec; \citealt{gaiagsp}) provides a good sample to search for metal-poor stars, but most Gaia GSP-Spec sources have $G < 13$. 

However, there exist (at least) two flaws in the Gaia photometric metallicites mentioned above. One is that the metallicities of some metal-rich stars were significantly under-estimated due to the effects of binary or phot\textunderscore bp\textunderscore rp\textunderscore excess\textunderscore factor\footnote{ Note that the binary effect and the phot\textunderscore bp\textunderscore rp\textunderscore excess\textunderscore factor effect show a similar behavior, as binaries tend to have a slightly larger phot\textunderscore bp\textunderscore rp\textunderscore excess\textunderscore factor.}, and mis-classified as metal-poor stars. 
Such a flaw can be largely avoided with the use of  additional criteria, as described below.
The other is that we failed to obtain metallicities for a large ($\sim$ 30 \%) fraction of VMPs because their Gaia colors lie beyond the
boundaries of the metallicity-dependent stellar locus of 
\cite{xu2022}; as a result, those stars were dropped in their catalog. 
We bring them back in this work because 
they are very likely VMPs, and we find that we can assign reasonably precise metallicities down to [Fe/H] = $-2.5$ through the use of additional criteria.

\section{Methodologies Employed} \label{sec:method}

Due to the relatively small number of bright VMP dwarf stars in the 
solar neighbourhood, we only focus on giants with $G < 15$ in this work. A relative error cut of parallax\_over\_error $>$ 2  is also applied to remove stars with likely large distance errors.

We have further adopted three independent criteria to select VMP candidates: the Gaia criterion, the WISE criterion, and the kinematic criterion.
The criteria can be used independently, or in combination, in order to select VMP samples with different efficiency.  Below we describe these three criteria.

\subsection{Gaia Criterion}

The Gaia criterion is very straightforward.
We simply select candidate giant stars with [Fe/H] $< - 2$ from \cite{xu2022}.
But it has two limitations. One is that it cannot
include stars with $|b| < 10^\circ$. The other is contamination 
from more metal-rich stars due to the effects of binary
or phot\textunderscore bp\textunderscore rp\textunderscore excess\textunderscore factor, as mentioned in 
Section\,2.

\subsection{WISE Criterion}
 
Following \cite{schlaufman2014}, we also use the WISE colors to
select VMP candidates. However, differing from \cite{schlaufman2014}, we have performed reddening corrections and use an updated criterion to improve the selection efficiency.
                        
\begin{figure}[htbp]
    \centering
    \includegraphics[width=8cm]{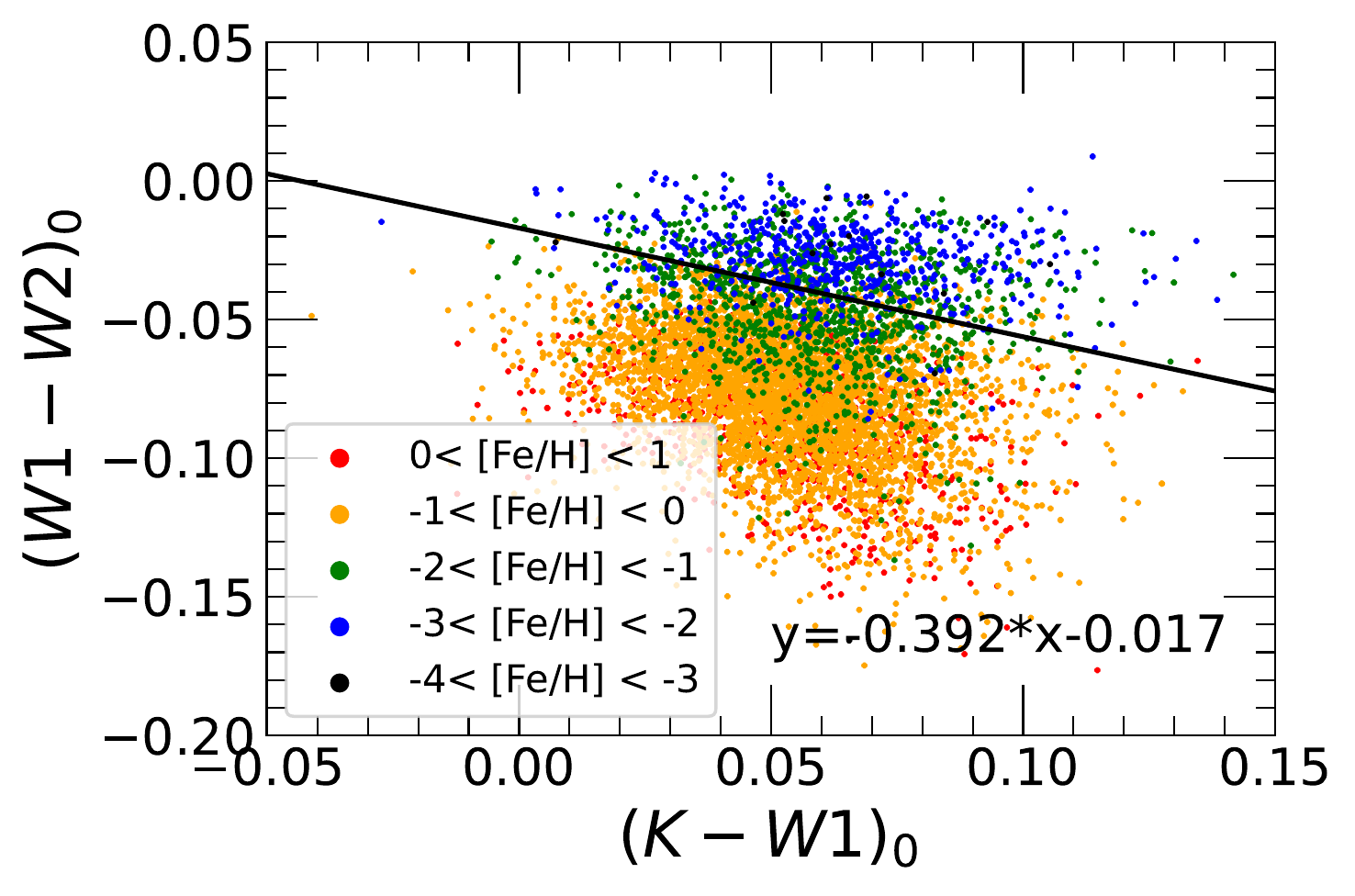}
    \caption{Distribution of stars in the $(W1 - W2)_0$ vs. $(K - W1)_0$ diagram. Colors indicate sources with different [Fe/H]. The black line is used to select candidate VMP stars with [Fe/H] $< - 2$.} 
    \label{fig:wise_figure}
\end{figure}

To determine the new WISE criterion, we use a selected sample of giant stars from LAMOST DR8 (\citealt{lamostdr8}) that have high-quality 2MASS and ALLWISE photometry with errors smaller than 0.025 mag and reddening values smaller than 0.01 mag.  
These stars are dereddened using the  \citeauthor*{SFD1998} (\citeyear{SFD1998}, hereafter SFD) dust reddening map. The reddening coefficients  
for the $K - W1$ and $W1-W2$ colors are empirically determined 
by Zhang et al. (2022, in preparation), which are 0.11 and 0.056, respectively.
With the sample above, by considering both the selection efficiency and completeness, we set the new criterion to be ${(W1 - W2)}_0 > -0.392 \times {(K - W1)}_0 - 0.017$, as shown in Figure \ref{fig:wise_figure}; it yields a success rate of 46$\%$ and a completeness of 84$\%$ for the sample used.

With the new criterion, we cross-match the Gaia EDR3, 2MASS, and ALLWISE catalogs to first select giant stars, then the subset of these that are candidate VMP stars.
For stars with $|b| \ge 10^\circ$, the giant stars are from \cite{xu2022}.
and the SFD reddening map is used. 
For stars with $|b| < 10^\circ$, the three-dimensional dust map from \cite{chen2019} is used to do the reddening correction, which provides $E(G - K_S)$, $E(BP - RP)$ and $E(H - K_S)$. The giant stars 
are selected as those with $M_{K_S} < -1.5\times (BP - RP)_0^2 +6.5\times (BP-RP)_0 - 3.8$, assuming that $A_{K_S} = 1.987 \times E(H - K_S)$, as given by \citet{yuan2013}. 
Note there is also a cut of $d < 6$ kpc, because of the distance limit of \cite{chen2019}. To avoid low success rates caused by large photometric errors, we also require that errors of the $Ks, W1$, and $W2$ bands are lower than 0.03, 0.025, and 0.025 mag, respectively. 
Note that the $W1-W2$ color can also be used to select M giant stars and estimate their metallicities (\citealt{Li_2016}).

\subsection{Kinematic Criteria}
As mentioned above, some metal-rich stars are mis-classified
as metal-poor stars with the Gaia photometric metallicities.
Most metal-rich stars are disk stars, while most VMP ones are halo stars. Since the majority of disk stars and halo stars have different
motions, we explore kinematics to exclude metal-rich stars from candidate Gaia VMP stars.                      

\begin{figure*}[htbp]
    \centering
    \includegraphics[width=16cm]{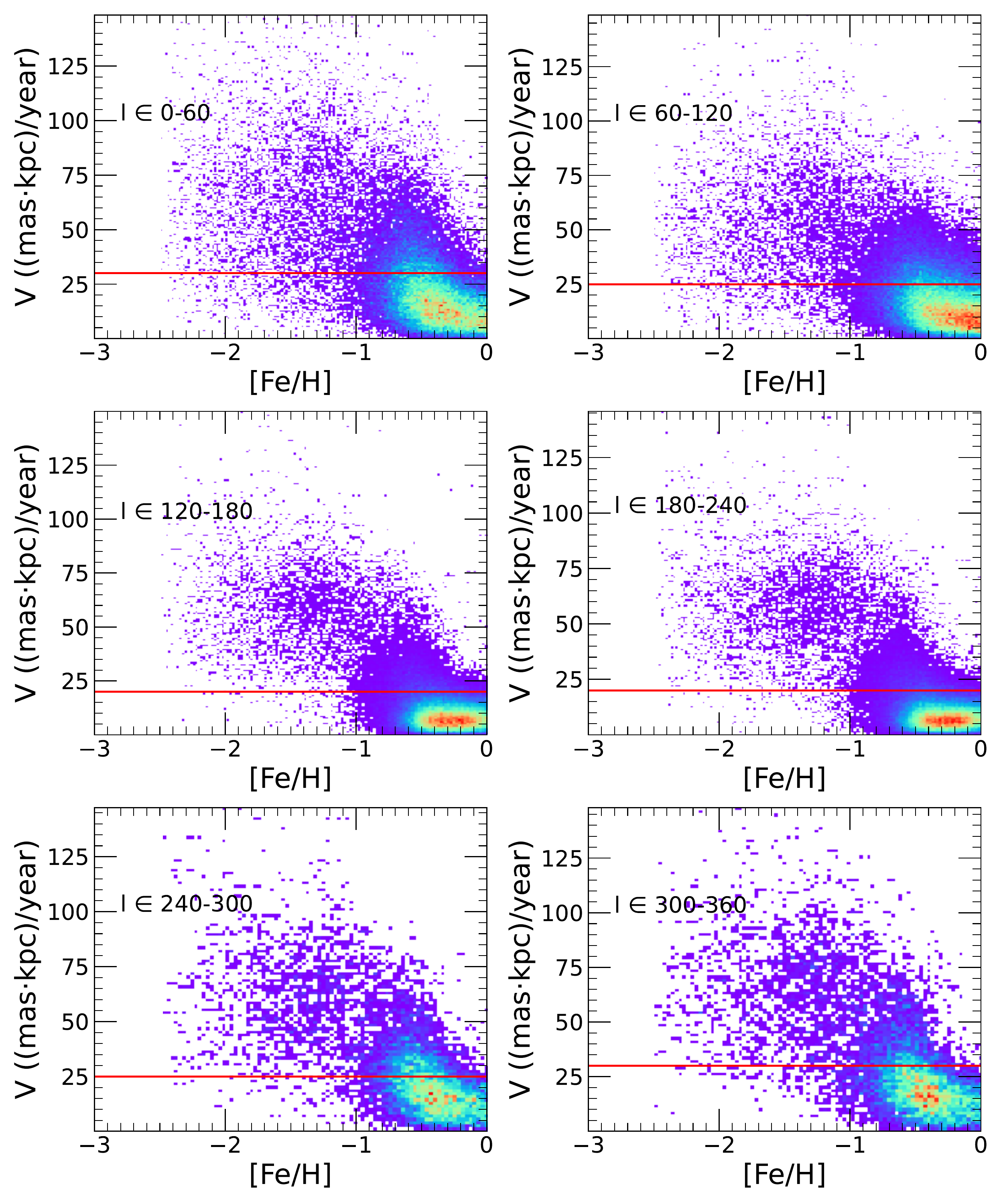}
    \caption{Tangential velocity, as a function of [Fe/H], for stars within different Galactic longitude ranges. The red lines are used to remove metal-rich disk stars. Note that the unit for
    tangential velocities is mas\,kpc\,year$^{-1}$, which can be converted to km\,s$^{-1}$ by multiplying by 4.74.}
    \label{fig:v_criteria}
\end{figure*}

Figure\,\ref{fig:v_criteria} shows the Gaia tangential velocities, as a function of [Fe/H], for stars over different Galactic longitude ranges. Note that the unit for tangential velocities is mas\,kpc\,year$^{-1}$, which can easily be converted to km\,s$^{-1}$ by multiplying by 4.74.  Here the stars are from the giant test sample of \cite{xu2022}. 
One can see that  most metal-rich ([Fe/H] $> -1$) stars have tangential velocities lower than 40 mas\,kpc\,year$^{-1}$,  but metal-poor stars ([Fe/H] $< -1$) have a much wider range, from 0 to 140 mas\,kpc\,year$^{-1}$.  Therefore, one can make a cut on tangential velocities to remove of the most metal-rich stars. The limits (red lines in Figure \ref{fig:v_criteria}) are different: 30 mas\,kpc\,year$^{-1}$ for sources within $- 60^\circ < l <60^\circ$, 20 mas\,kpc\,year$^{-1}$ for sources with $120^\circ < l < 240^\circ $, and 25 mas\,kpc\,year$^{-1}$ for sources with $60^\circ < l < 120^\circ $ or $240^\circ < l < 300^\circ$. Such a kinematic criterion can be used together with the Gaia and WISE criteria to select cleaner VMP samples, albeit with the introduction of a kinematic bias. Note that a small number of VMP stars with disk-like orbits are likely excluded by this criterion.

\subsection{Different Samples}

There are three criteria in our work, as described above. For sources with $|b| > 10^\circ$, we define a Gold sample, two Silver samples, and two Bronze samples, corresponding to the confidence we place in the selection of VMP stars. The Gold sample satisfies all three criteria. The Silver GK sample satisfies the Gaia and Kinematic criteria, and the Silver GW sample satisfies the Gaia and WISE criteria.
The Bronze G sample satisfies the Gaia criterion only. 
The Bronze WK sample satisfies the WISE and Kinematic criteria. For sources with $|b| < 10^\circ$, we use 
the WISE and kinematic criteria to select VMP stars, and refer to 
them as the Low $b$ sample.  The samples and their selection criteria are summarized in Table \ref{table1}. 

By cross-matching with results derived from medium-resolution ($R \sim$ 1800) from LAMOST DR8, we establish that the success rate for identifying VMP stars is 60.1$\%$ for the Gold sample, 39.2$\%$ for the Silver GW sample, 41.3$\%$ for the Silver GK sample, 15.4$\%$ for the Bronze G sample, 31.7$\%$ for the Bronze WK sample, and 16.6$\%$ for the Low $b$ sample, respectively. 

\setlength{\tabcolsep}{3mm}{
\begin{table}[htbp]
\footnotesize
\centering
\caption{Different samples and their selection criteria. }
\begin{tabular}{l|c|c|c} 

          & Kinematic  & WISE       & Gaia     \\
\hline
$\rm Gold^a$      & \checkmark & \checkmark & \checkmark\\
$\rm Silver\, GW^a$  &            & \checkmark & \checkmark\\
$\rm Silver\, GK^a$ & \checkmark &            & \checkmark\\
$\rm Bronze\, G^a $ &            &            & \checkmark\\
$\rm Bronze\, WK^a$ & \checkmark & \checkmark &           \\
$\rm Low\, $b$^b    $ & \checkmark & \checkmark &      \\ \hline
\end{tabular}
\label{table1}
$^a$ For sources with $|b| > 10^\circ$. 
$^b$ for sources with $|b| < 10^\circ$.
\end{table}}

\section{Results}\label{sec:result}

After applying the criteria described above,
for sources with $|b| > 10^\circ$, we have identified
24,304 Gold sample stars, 40,157 Silver GW sample stars, 
120,452 Silver GK sample stars, 
291,690 Bronze G sample stars, and 68,526 Bronze WK sample stars.
A total of 4,645 stars with $|b| < 10^\circ$ is also collected in the Low $b$ sample.
All of the samples are publicly available\footnote{\url{https://doi.org/10.12149/101160}}.
The columns for the five samples with sources of $|b|>10^\circ$ are the same as listed in Table \ref{table:discription}. 
The columns for the Low $b$ sample are listed in Table \ref{table:discription_low_b}. 
In this section, we describe our six samples and discuss their relative success rates.

\subsection{Properties of the Samples}

\begin{figure*}[htbp]
    \centering
    \includegraphics[width=14cm]{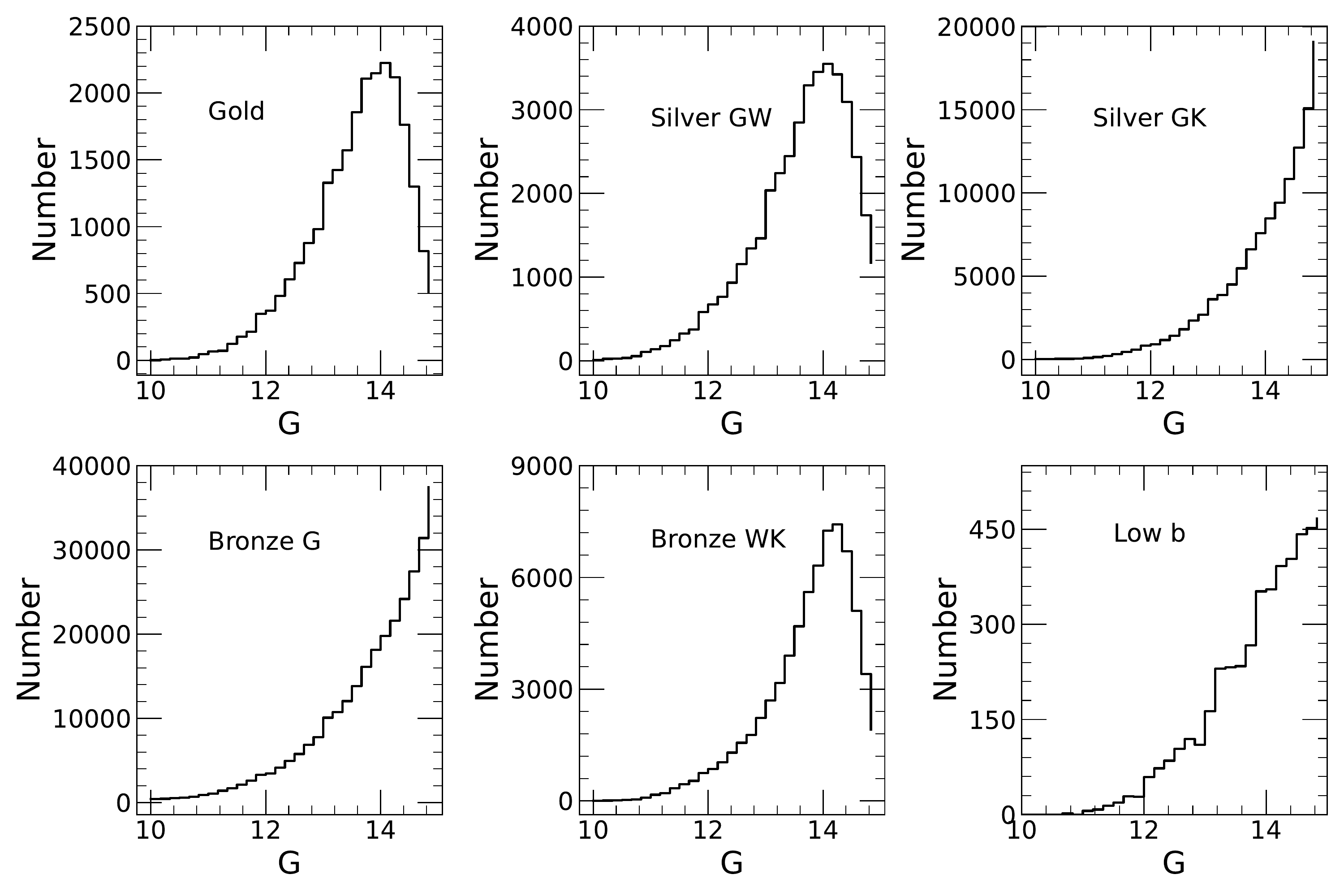}
    \caption{Histograms  of the $G$ magnitude for our six samples.}
    \label{fig:G_distribution}
\end{figure*}

The $G$-magnitude distributions for the six samples are shown in Figure \ref{fig:G_distribution}. 
All of the candidate stars are brighter than 15 in the $G$-band, which are relatively easy to access with high-resolution follow-up spectroscopic observations. 
There is a peak around $G \sim 14$ in the Gold, Silver GW, and Bronze WK samples, due to the cut on photometric errors in the WISE criterion. 

\begin{figure*}[htbp]
\centering
\subfigure{
\includegraphics[width=8.5cm]{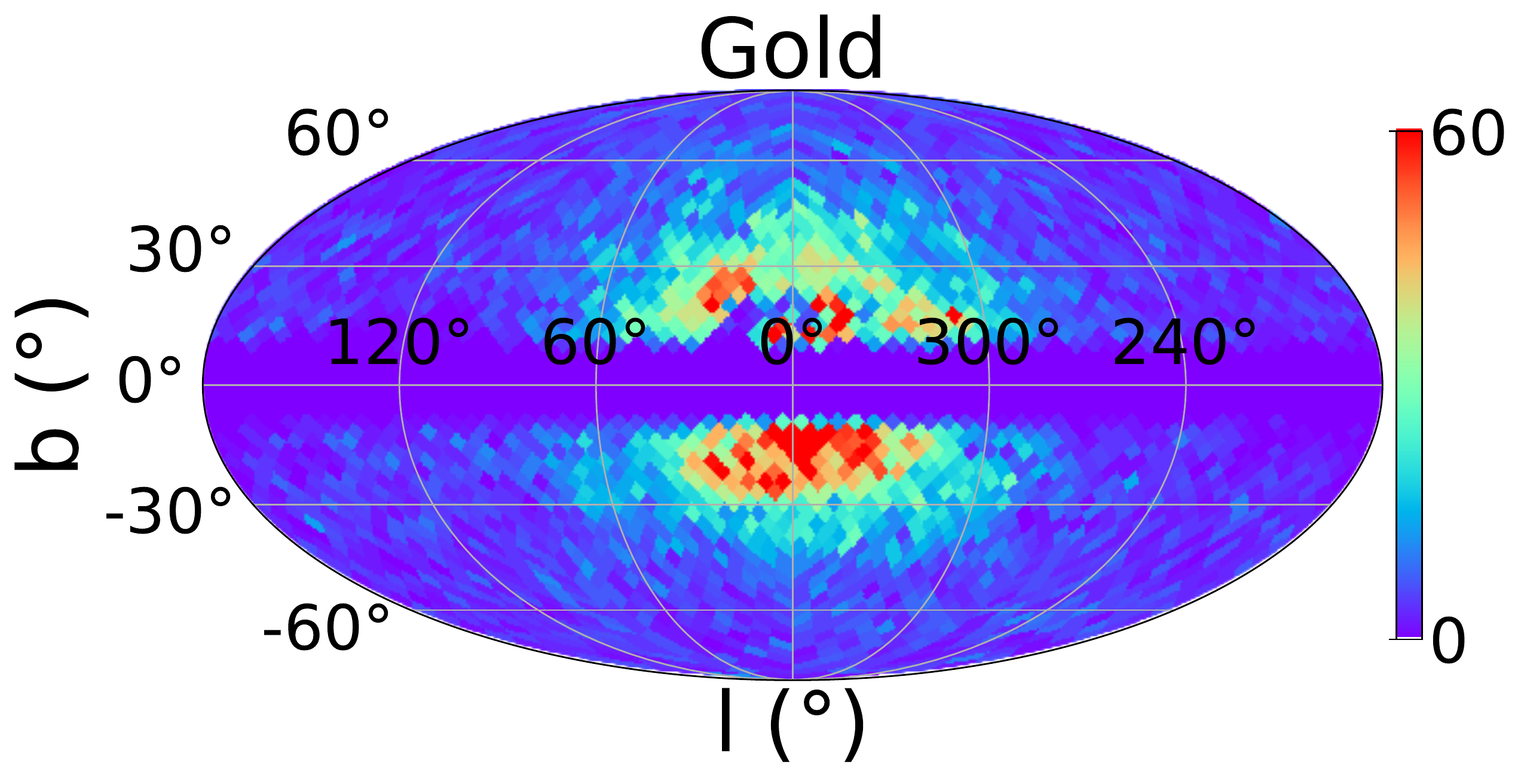}
}
\subfigure{
\includegraphics[width=8.5cm]{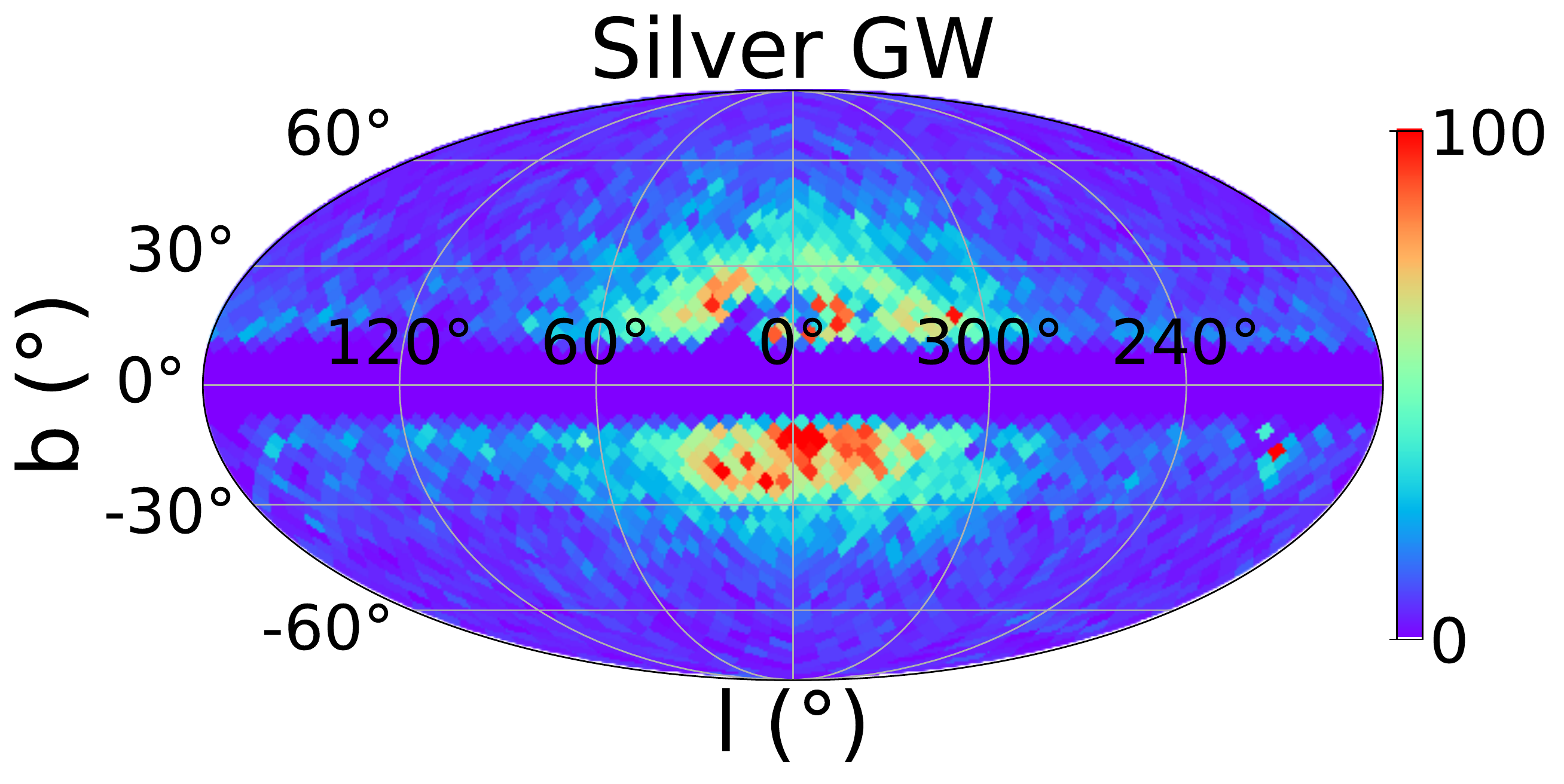}
}

\subfigure{
\includegraphics[width=8.5cm]{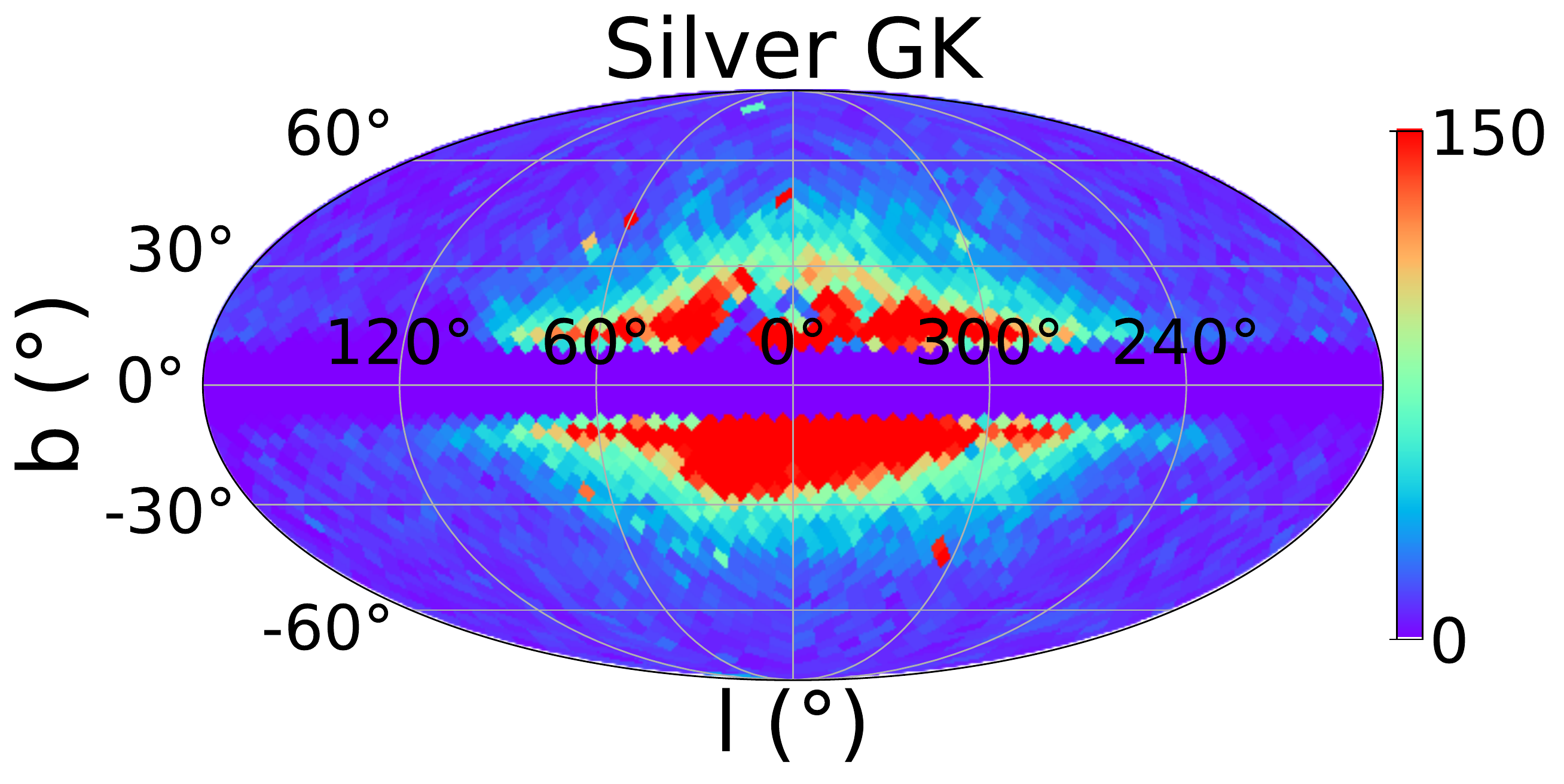}
}\subfigure{
\includegraphics[width=8.5cm]{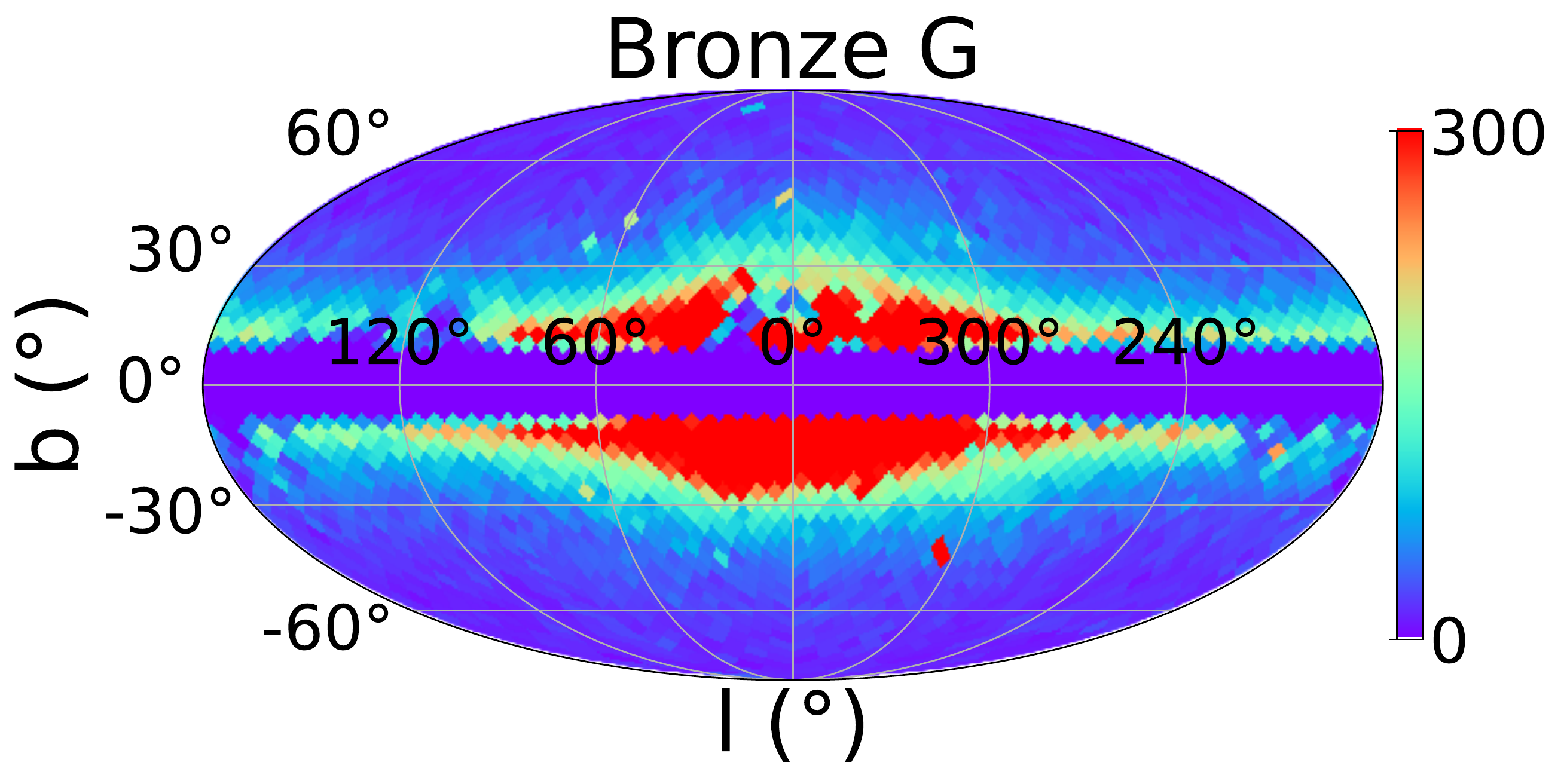}
}

\subfigure{
\includegraphics[width=8.5cm]{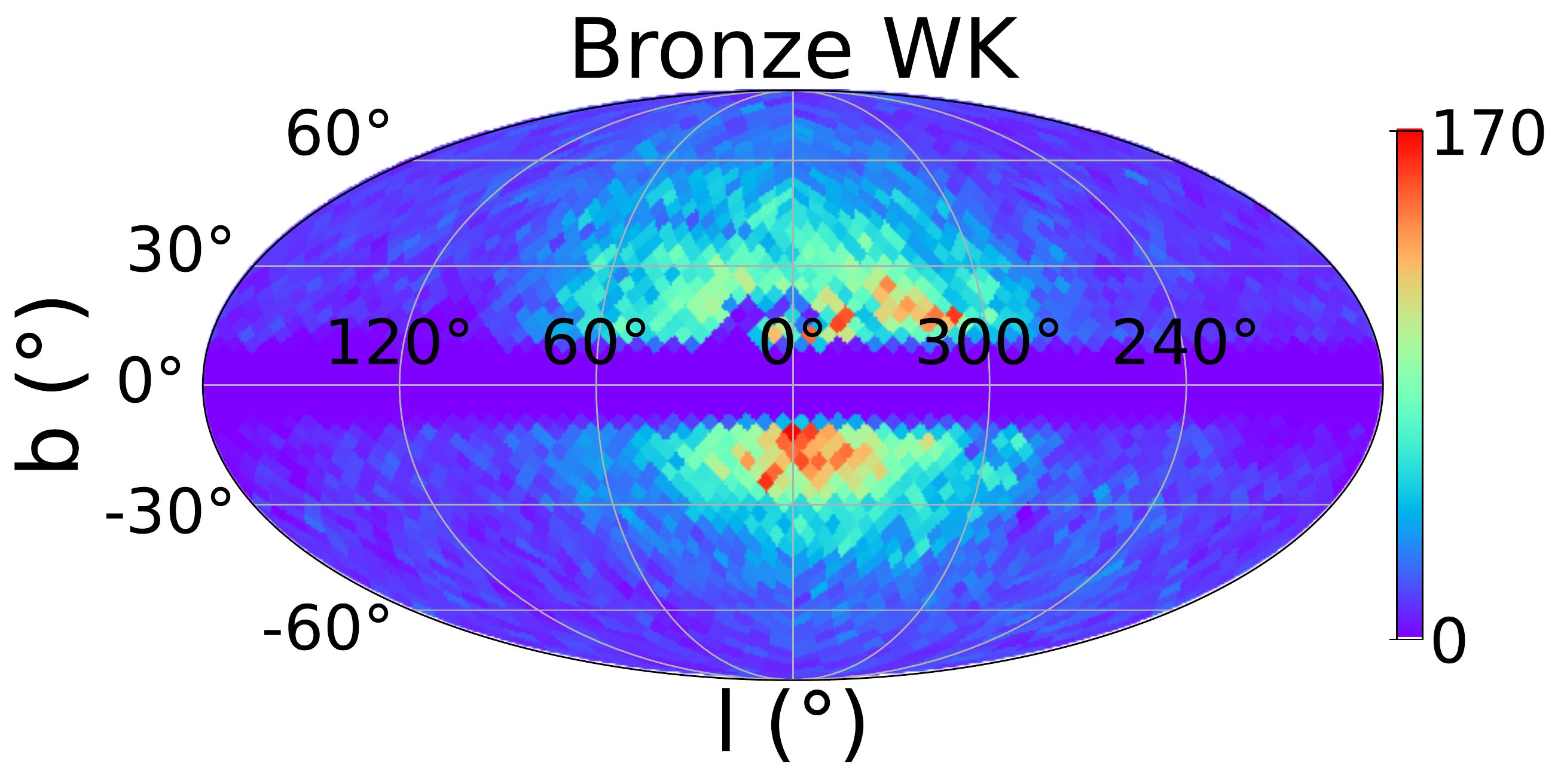}
}\subfigure{
\includegraphics[width=8.5cm]{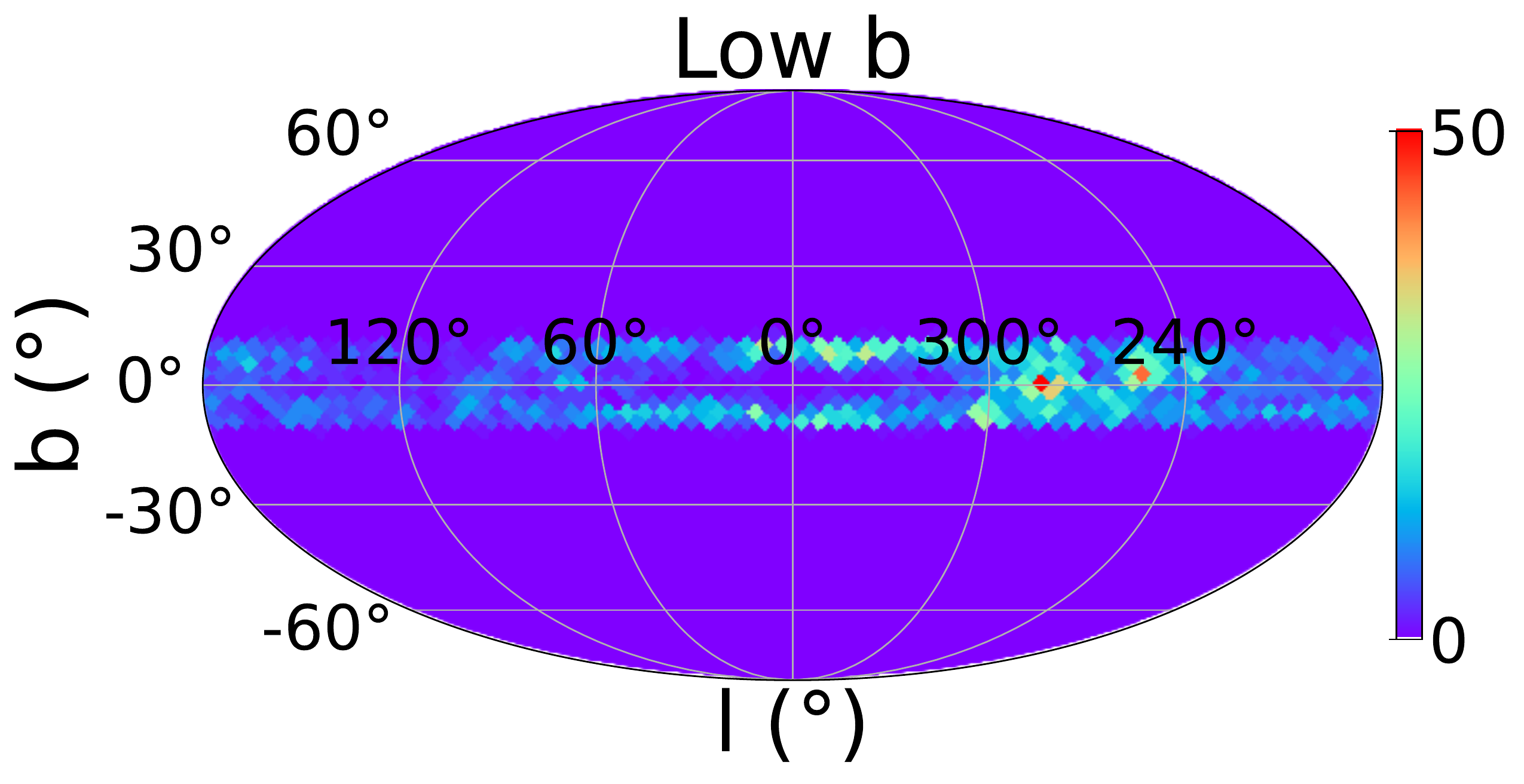}
}
\caption{Spatial distributions of the six samples in the Galactic coordinate system. Colors indicate number densities on a linear scale.}
\label{fig:space_ditribution}
\end{figure*}

Spatial distributions of the six samples in the Galactic coordinate system are shown in Figure \ref{fig:space_ditribution}. The sources are roughly evenly distributed over high Galactic latitude areas, and slightly increase toward the Galactic center direction. 
For the Bronze G sample, there is a strong over-density 
in the dense Galactic center and disk regions, mainly because there are more metal-rich stars mis-classified as VMP stars. Such an over-density is significantly reduced by application of the Wise and Kinematic criteria. For the Gold sample, there is still a weak over-density in the Galactic center direction, which is likely real
to some extent, because we expect to have more VMP stars there compared 
to those near the Sun's position in the Galaxy.

\setlength{\tabcolsep}{4.7mm}{
\begin{table*}[htbp]
\footnotesize
\centering
\caption{Description of the Gold, Silver GW, Silver GK, Bronze G, and Bronze WK Samples}
\begin{tabular}{lll}
\hline
\hline
Field & Description & Unit\\
\hline
source\_id & Unique source identifier for EDR3 (unique with a particular Data Release) & --\\
ra & Right ascension & deg\\
dec & Declination & deg\\
parallax & Parallax & mas\\
parallax\_error & Standard error of parallax & mas\\
pmra & Proper motion in right ascension direction & mas/year\\
pmra\_error & Standard error of proper motion in right ascension direction & mas/year\\
pmdec & Proper motion in declination direction & mas/year\\
pmdec\_error & Standard error of proper motion in declination direction & mas/year\\
ruwe & Renormalised unit weight error & --\\
phot\_g\_mean\_flux\_over\_error &  $G$-band mean flux divided by its error & --\\
phot\_g\_mean\_mag & $G$-band mean magnitude & --\\
phot\_bp\_mean\_flux\_over\_error & $BP$-band mean flux divided by its error & --\\
phot\_bp\_mean\_mag & Integrated $BP$-band mean magnitude & --\\
phot\_rp\_mean\_flux\_over\_error & $RP$-band mean flux divided by its error & --\\
phot\_rp\_mean\_mag & Integrated $RP$-band mean magnitude & --\\
phot\_bp\_rp\_excess\_factor & $BP/RP$ excess factor & --\\
l & Galactic longitude & deg\\
b & Galactic latitude & deg\\
ebv & Value of E (B $-$ V ) from from the extinction map of SFD98 & --\\
correct\_bp\_rp & Intrinsic $BP - RP$ color after color correction of \cite{Niu2021EDR3} & --\\
correct\_bp\_g & Intrinsic $BP - G$ color after color correction of \cite{Niu2021EDR3} & --\\
correct\_g\_rp & Intrinsic $G - RP$ color after color correction of \cite{Niu2021EDR3} & --\\
FeH\_Gaia & Photometric metallicity & -- \\
FeH\_Gaia\_error & Formal error of FeH\_Gaia  & dex\\
W1mag & $W1$ magnitude & --\\
W2mag & $W2$ magnitude & --\\
Jmag & $J$ magnitude & --\\
Hmag & $H$ magnitude & --\\
Kmag & $K_S$ magnitude & --\\
e\_W1mag & Error of $W1$ magnitude  & mag\\
e\_W2mag & Error of $W2$ magnitude & mag\\
e\_Jmag & Error of $J$ magnitude & mag\\
e\_Hmag & Error of $H$ magnitude & mag\\
e\_Kmag & Error of $K_S$ magnitude & mag\\
parallax\_corrected & Parallax corrected by \cite{lindegren2021} & mas\\
radial\_velocity & radial velocity & km\,s$^{-1}$\\
radial\_velocity\_error & radial velocity error & km\,s$^{-1}$\\
\hline
\end{tabular}
\label{table:discription}
\end{table*}}

\setlength{\tabcolsep}{4.7mm}{
\begin{table*}[htbp]
\footnotesize
\centering
\caption{Description of the Low $b$ Sample}
\begin{tabular}{lll}
\hline
\hline
Field & Description & Unit\\
\hline
source\_id & Unique source identifier for EDR3 (unique with a particular Data Release) & --\\
ra & Right ascension & deg\\
dec & Declination & deg\\
parallax & Parallax & mas\\
parallax\_error & Standard error of parallax & mas\\
pmra & Proper motion in right ascension direction & mas/year\\
pmra\_error & Standard error of proper motion in right ascension direction & mas/year\\
pmdec & Proper motion in declination direction & mas/year\\
pmdec\_error & Standard error of proper motion in declination direction & mas/year\\
ruwe & Renormalised unit weight error & --\\
phot\_g\_mean\_flux\_over\_error &  $G$-band mean flux divided by its error & --\\
phot\_g\_mean\_mag & $G$-band mean magnitude & --\\
phot\_bp\_mean\_flux\_over\_error & $BP$-band mean flux divided by its error & --\\
phot\_bp\_mean\_mag & Integrated $BP$-band mean magnitude & --\\
phot\_rp\_mean\_flux\_over\_error & $RP$-band mean flux divided by its error & --\\
phot\_rp\_mean\_mag & Integrated $RP$-band mean magnitude & --\\
phot\_bp\_rp\_excess\_factor & $BP/RP$ excess factor & --\\
l & Galactic longitude & deg\\
b & Galactic latitude & deg\\
correct\_bp\_rp & Intrinsic $BP - RP$ color after color correction of \cite{Niu2021EDR3}  & --\\
 & and reddening correction of  \citet{chen2019}  & \\
W1mag & $W1$ magnitude & --\\
W2mag & $W2$ magnitude & --\\
Jmag & $J$ magnitude & --\\
Hmag & $H$ magnitude & --\\
Kmag & $K_S$ magnitude & --\\
M\_K\_S & Absolute magnitude in the $K_S$ band\\
e\_W1mag & Error of $W1$ magnitude & mag\\
e\_W2mag & Error of $W2$ magnitude & mag\\
e\_Jmag & Error of $J$ magnitude & mag\\
e\_Hmag & Error of $H$ magnitude & mag\\
e\_Kmag & Error of $K_S$ magnitude & mag\\
parallax\_corrected & Parallax corrected by \cite{lindegren2021} & mas\\
radial\_velocity & radial velocity & km\,s$^{-1}$\\
radial\_velocity\_error & radial velocity error & km\,s$^{-1}$\\
\hline
\end{tabular}
\label{table:discription_low_b}
\end{table*}}

\subsection{Success Rates of the Samples}

We cross-match our six samples with LAMOST DR8 (\citealt{lamostdr8}) in order to test the success rates of our various criteria. LAMOST DR8 has provided values of stellar atmospheric parameters for 5.16 million unique stars, including reliable estimates of [Fe/H] down to $\sim -3.5$ (see Figure A4 of \citealt{lamostdr8}). 

\begin{figure*}[htbp]
    \centering
    \includegraphics[width=16cm]{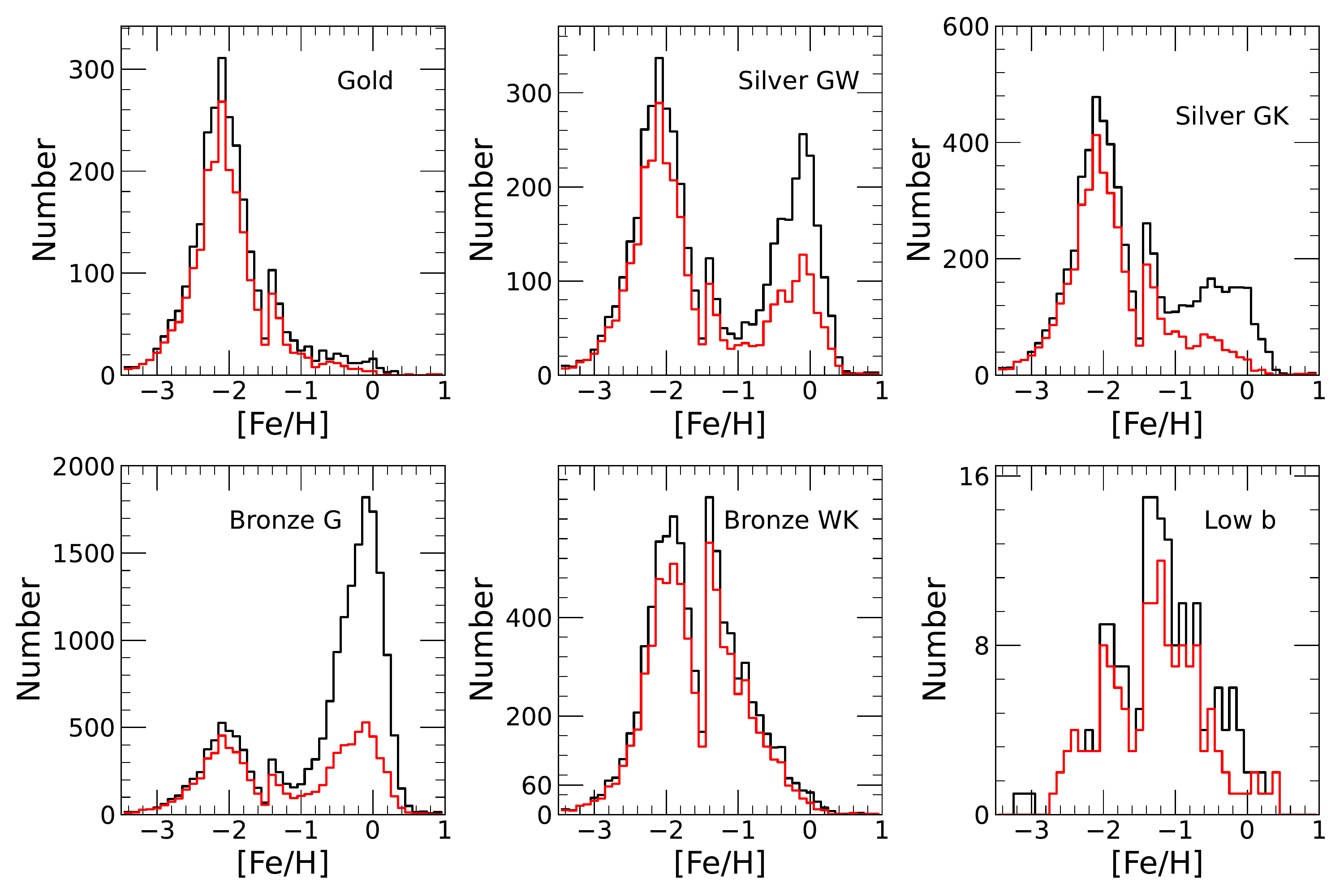}
    \caption{[Fe/H] distributions of the sources in common between the six samples and LAMOST DR8. The black lines are for the default samples and the red lines are for samples applying the cut on $RUWE$.}
    \label{fig:success}
\end{figure*}

There are 2764, 4719, 6153, 18353, 8377, and 193 stars found in common 
for the Gold, Silver GW, Silver GK, Bronze G, Bronze WK, and Low $b$ samples, respectively, after a 
signal-to-noise cut in the $g$-band ($S/N_g > 20$). Their [Fe/H] distributions are plotted in black in Figure \ref{fig:success}.
The Bronze G, Silver GW, and Silver GK samples show two clear peaks, one for the VMP stars, the other for the metal-rich stars. The metal-rich peak is most prominent in the Bronze G sample, as explained in Section\,2. 
The metal-rich peak becomes weaker and weaker in the Silver GW and GK samples, and disappears in the Gold sample. The result suggests that the WISE and Kinematic criterion, particularly the latter,  work well in removing contamination from metal-rich stars. 
There are quite large fractions of metal-poor stars ($-2 <$ [Fe/H] $< -1$) in the Bronze WK and Low $b$ samples, both of which have used the WISE criterion. This is perhaps not surprising, given the relatively large errors in the WISE colors.
Note that there is a dip at [Fe/H] $\sim -1.5$ in all samples, most prominent in the Bronze WK sample. This arises because, in LAMOST DR8, the metallicities for stars with [Fe/H] $> -1.5$ and $< -1.5$ are measured by two different pipelines (see Section\,5 of \citealt{lamostdr8}).

For a given sample, we define its success rate for selecting stars 
with [Fe/H] lower than a given limit as the ratio of DR8 stars in common
with [Fe/H] lower than the limit, relative to the total number of stars in common. The results are listed in Table \ref{table:success}.
For the Gold sample, the success rate is 93.0\%, 83.1\%, 60.1\%, and 16.2\% for stars with [Fe/H] lower than $-1$, $-1.5$, $-2$, and $-2.5$, respectively. 
For the two Silver samples, the success rate is somewhat lower, 
about 40\% for stars with [Fe/H] $< -2$. 
It decreases to 31.8\% for the Bronze WK sample, 
15.4\% for the Bronze G sample, and 16.6\% for 
the Low $b$ sample, respectively.

The Bronze G sample shows the lowest success rate in all cases, 
due to the much larger number of metal-rich binary stars compared to VMP stars. To increase its success rate,
we use the Renormalized Unit Weight Error 
(RUWE; \citealt{lindegren2021b}) to exclude binaries with poor astrometric solutions. 
The sources with larger RUWE values are more likely binaries. 
We find that an empirical cut of $RUWE < 1.1$ (red lines in
Figure \ref{fig:success}) can well-remove 
metal-rich binaries for the Gaia criterion\footnote{ 
Note the cut on RUWE is only used to remove binaries, regardless of their metallicites.  However, as most binaries are metal-rich in terms of numbers, it seems that the cut preferentially removes metal-rich binaries.}.
With this simple additional criterion, 
the success rate of VMP stars increases to 63.1$\%$ for the Gold sample, about 50$\%$ for the Silver samples, and 29.3$\%$ for the Bronze G sample, as listed in
Table \ref{table:success}. 

\setlength{\tabcolsep}{2.5mm}{
\begin{table*}[htbp]
\footnotesize
\centering 
\caption{Success Rates of the Six Samples by Comparing with LAMOST DR8}
\begin{tabular}{l|c|c|c|c|c|c|c} 
Sample & Total No.  & No. of common sources & [Fe/H] $<-3$  & [Fe/H] $<-2.5$ & [Fe/H] $<-2$ & [Fe/H] $<-1.5$ & [Fe/H] $<-1$\\
\hline
$\rm Gold^a$ &  24,304 & 2,764 & 2.9\% & 16.2\% & 60.1\% & 83.1\% & 93.0\% \\
$\rm Gold^b$ &  20,436 & 2,139 & 3.3\% & 17.4\% & 63.1\% & 86.2\% & 95.7\% \\

\hline
$\rm Silver GW^a$ & 40,157 & 4,719 & 1.9\% & 10.9\% & 39.2\% & 54.6\% & 61.8\% \\
$\rm Silver GW^b$ & 30,492 & 3,272 & 2.4\% & 13.3\% & 46.9\% & 64.8\% & 72.7\% \\

\hline
$\rm Silver GK^a$ & 120,452 & 6,153 & 2.1\% & 11.1\% & 41.3\% & 60.0\% & 73.3\% \\
$\rm Silver GK^b$ & 91,952  & 4,168 & 2.8\% & 14.3\% & 51.6\% & 73.4\% & 87.4\% \\

\hline
$\rm Bronze G^a$ & 291,690 & 18,353 & 0.8\% & 4.2\%  & 15.4\% & 22.4\% & 28.2\% \\
$\rm Bronze G^b$ & 173,544 & 8,185  & 1.6\% & 9.3\%  & 29.3\% & 41.8\% & 50.6\% \\

\hline
$\rm Bronze WK^a$ & 69,526 & 8,377 & 1.3\% & 6.8\%  & 31.8\% & 56.0\% & 82.5\% \\
$\rm Bronze WK^b$ & 60,896 & 7,065 & 1.4\% & 6.9\%  & 31.7\% & 56.0\% & 83.2\% \\
\hline
$\rm Low $b$^a  $ & 4,645 & 193 & 1.5\% & 4.7\%  & 16.6\% & 32.6\% & 66.3\% \\
$\rm Low $b$ ^b $ & 3,910 & 144 & 0.0\% & 4.2\%  & 18.8\% & 36.1\% & 68.8\% \\\hline

\end{tabular}
$^a$ Default samples.
$^b$ Samples after applying the additional cut of $RUWE < 1.1$.

\label{table:success}
\end{table*}
}

\begin{figure*}[htbp]
    \centering
    \includegraphics[width=16cm]{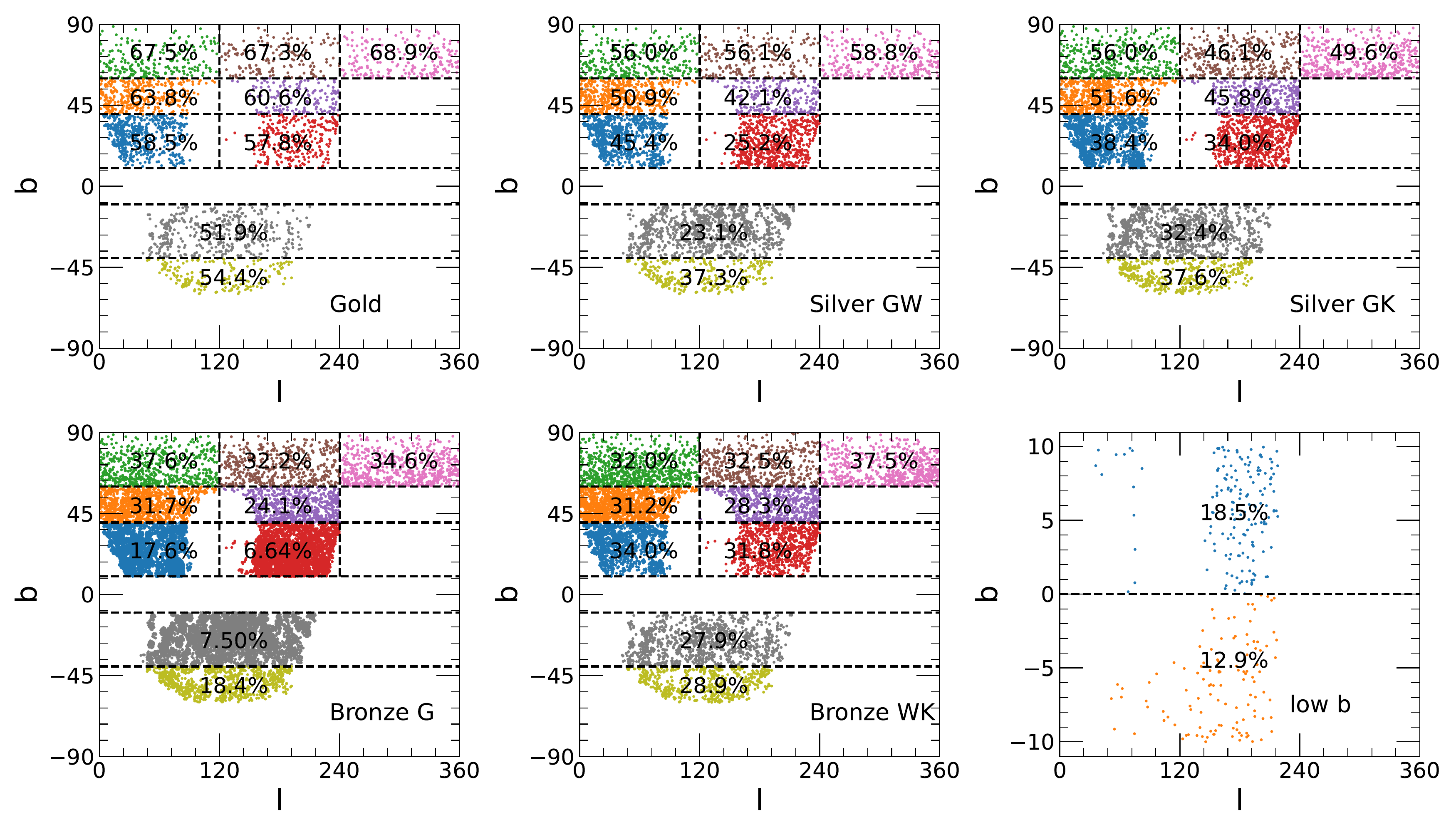}
    \caption{Spatial variations of the success rate for the six samples based on their sources in common with LAMOST DR8.}
    \label{fig:test_lamost}
\end{figure*}

Figure \ref{fig:test_lamost} plots the success rate of VMP stars identified in different sky areas. 
The success rate is relatively larger in the higher Galactic latitude regions due to less contamination. For samples using the Gaia criterion, 
the success rate in the Southern Galactic Hemisphere is lower than that in the Northern Galactic Hemisphere, likely due to the different systematics in the SFD reddening map (\citealt{Schlafly_2010, Sun2022}). Systematic errors in the SFD map result in systematic errors in the Gaia metallicities. 

We also cross-match our six samples with SEGUE DR12 (\citealt{sdssdr12}) in order to test the success rates of our various criteria. The numbers of stars in common are limited, due to the bright limit for SDSS ($g \sim 14$) -- only 49 stars for the Gold sample and 323 stars for the Bronze G sample. We find that the overall success rates are slightly better than those with LAMOST DR8, because most sources from SEGUE DR12 have $|b| > 20^\circ$, where higher success rates are found in the test with LAMOST DR8. 

\begin{figure}[htbp]
    \centering
    \includegraphics[width=8cm]{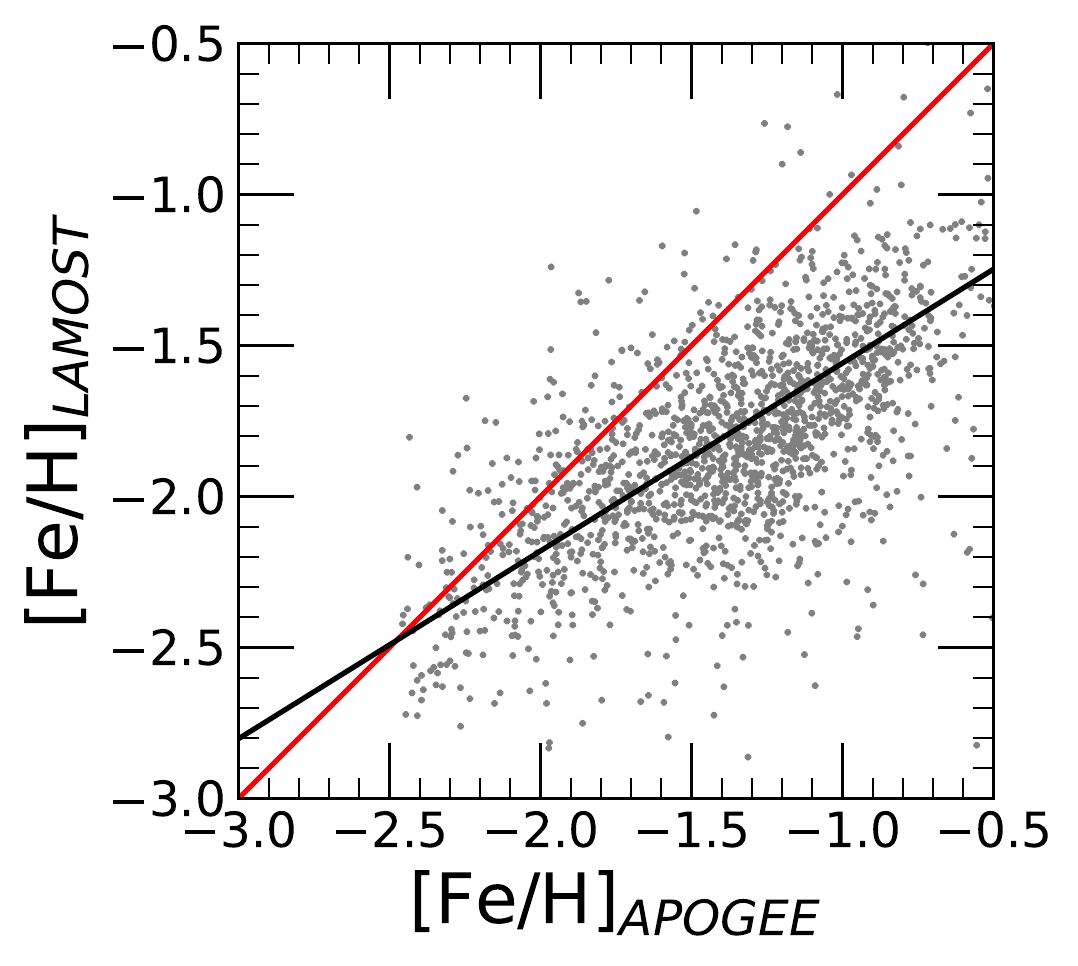}
    \caption{[Fe/H]$_{LAMOST}$ vs. Fe/H]$_{APOGEE}$. The red line is the one-to-one line. The black line is the linear fitting result: [Fe/H]$_{LAMOST}$ = 0.62 $\times$ [Fe/H]$_{APOGEE} - 0.94$.  We use it to convert Fe/H]$_{APOGEE}$ to [Fe/H]$_{LAMOST}$ (see text). }
    \label{fig:lamost_apogee}
\end{figure}

To analyse the success rate over the full sky, metallicities from the Apache Point Observatory Galactic Evolution Experiment(APOGEE; \citealt{apogee}) Data Release 17 (\citealt{apogeedr17}) are also used. 
There is an offset between the LAMOST and APOGEE metallicities, 
so we cross-match LAMOST DR8 giants of $S/N_g > 20$ and $T_{eff} > 4400$ K 
and those of APOGEE DR17 with $S/N > 100$ to correct for this.
Figure \ref{fig:lamost_apogee} plots [Fe/H] $_{LAMOST}$ vs. Fe/H] $_{APOGEE}$. The linear fitting 
result of [Fe/H] $_{LAMOST}$ = 0.62 $\times$ Fe/H] $_{APOGEE}$ $-$ 0.94 is over-plotted as the black line. The APOGEE metallicities are converted into 
LAMOST metallicities using the above relation. 

\begin{figure*}[htbp]
    \centering
    \includegraphics[width=16cm]{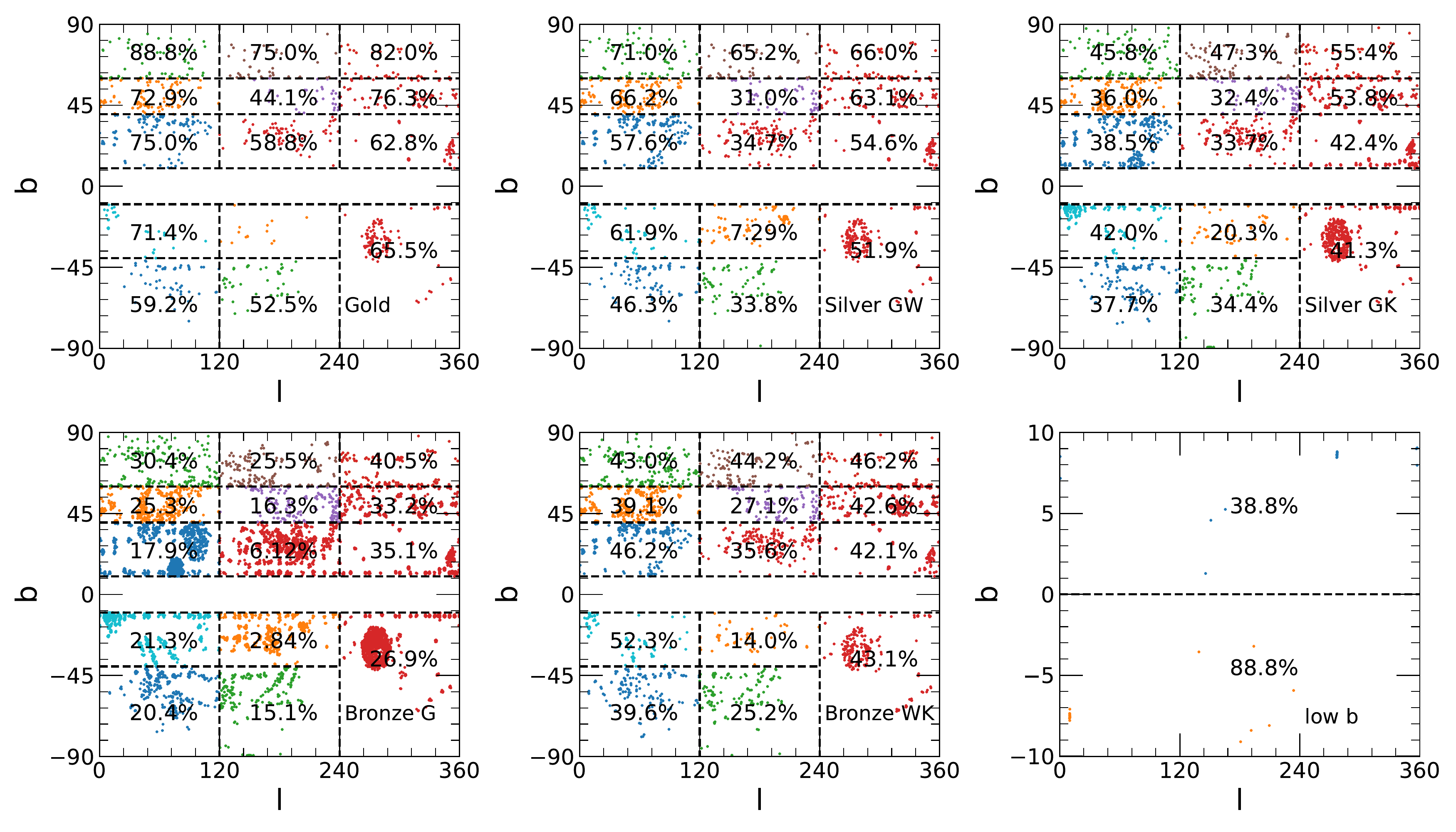}
    \caption{Same as Figure\,\ref{fig:test_lamost} but using the APOGEE DR17 stars. }
    \label{fig:test_apogee}
\end{figure*}

We cross-match our six samples with APOGEE DR17, and only sources with $S/N > 50$ are used in subsequent studies. The result is plotted in Figure \ref{fig:test_apogee}. The trend is similar to Figure\,\ref{fig:test_lamost}. Note the much lower success rate in 
the region $120^\circ < l < 240^\circ$ and $- 40^\circ < b <- 10^\circ$, due to systematics in the SFD map.

\section{Summary}\label{sec:con}

We have used three independent criteria, including 
the Gaia criterion, the WISE criterion, and the Kinematic criterion, to identify bright VMP giant candidate stars. 
The Gaia criterion is based on the photometric metallicities 
from \cite{xu2022}. The WISE criterion takes advantage of the lack of 
molecular absorption near 4.6 microns for VMP stars (Figure\,\ref{fig:wise_figure}). The Kinematic criterion relies on the generally much higher tangential velocities of VMP stars compared 
to disk stars (Figure\,\ref{fig:v_criteria}).

With different combinations of these criteria, we have collected 
six samples with $10 < G < 15$ (Table\,1): one Gold sample, two
Silver samples, two Bronze samples, and one Low $b$ sample.
The Gold, Silver GW and GK, Bronze G and WK samples contain
24,304, 40,157, 120,452, 291,690, and 68,526 VMP candidates with
$|b| > 10^\circ$, respectively. 
The Low $b$ sample contains 4,645 candidates with $|b| < 10^\circ$.
By cross-matching with the LAMOST DR8, the success rate for VMP stars is 60.1$\%$ for the Gold sample, 39.2$\%$ for the Silver GW sample, 41.3$\%$ for the Silver GK sample, 15.4$\%$ for the Bronze G sample, 31.7$\%$ for the Bronze WK sample, and 16.6$\%$ for the Low $b$ sample, respectively. The success rate varies with spatial 
position; this trend is further confirmed with APOGEE DR17.
If needed, a simple additional strict cut of $RUWE < 1.1$ can further increase these success rates, to 63.1$\%$ for the Gold sample, about 50$\%$ for the Silver samples, and 29.3$\%$ for the Bronze G sample.  Using 3D velocities, when available, rather than tangential velocities alone, will also increase the success rate of the Silver GK from 41.3$\%$ to 46.0$\%$ and Bronze WK samples from 31.8$\%$ to 38.5$\%$.

Our samples provide valuable candidates for high-resolution follow-up spectroscopic observations to find and study the most metal-poor stars. The samples are also useful in studies of the Galactic halo. 

With the latest release of Gaia DR3 BP/RP spectra, one expects to identify VMP stars more easily and accurately than from Gaia colors alone. Such explorations will be carried out in the near future.

\begin{acknowledgments}

This work is supported by the National Natural Science Foundation of China through the projects NSFC 12222301, 12173007, 11603002, 11933004, National Key Basic R \& D Program of China via 2019YFA0405500, Beijing Normal University grant No. 310232102. T.C.B. acknowledges partial support from grant PHY 14-30152,
Physics Frontier Center/JINA Center for the Evolution
of the Elements (JINA-CEE), awarded by the US National
Science Foundation. His participation in this work was initiated
by conversations that took place during a visit to China
in 2019, supported by a PIFI Distinguished Scientist award
from the Chinese Academy of Science. We acknowledge the science research grants from the China Manned Space Project with NO. CMS-CSST-2021-A08 and CMS-CSST-2021-A09. 

This work has made use of data from the European Space Agency (ESA) mission {\it Gaia} (\url{https://www.cosmos.esa.int/gaia}), processed by the Gaia Data Processing and Analysis Consortium (DPAC, \url{https:// www.cosmos.esa.int/web/gaia/dpac/ consortium}).
Funding for the DPAC has been provided by national institutions, in particular the institutions participating in the Gaia Multilateral Agreement. 
Guoshoujing Telescope (the Large Sky Area Multi-Object Fiber Spectroscopic Telescope LAMOST) is a National Major Scientific Project built by the Chinese Academy of Sciences. 
Funding for the project has been provided by the National Development and Reform Commission. LAMOST is operated and managed by the National Astronomical Observatories, Chinese Academy of Sciences.

\end{acknowledgments}

\appendix

\newcounter{Afigure}
\setcounter{Afigure}{1}
\renewcommand{\thefigure}{A\arabic{Afigure}}

\section{comparison of [Fe/H] with other catalogs}\label{sec:appendix}

\begin{figure*}[htbp]
\centering
\subfigure{
\includegraphics[width=7cm]{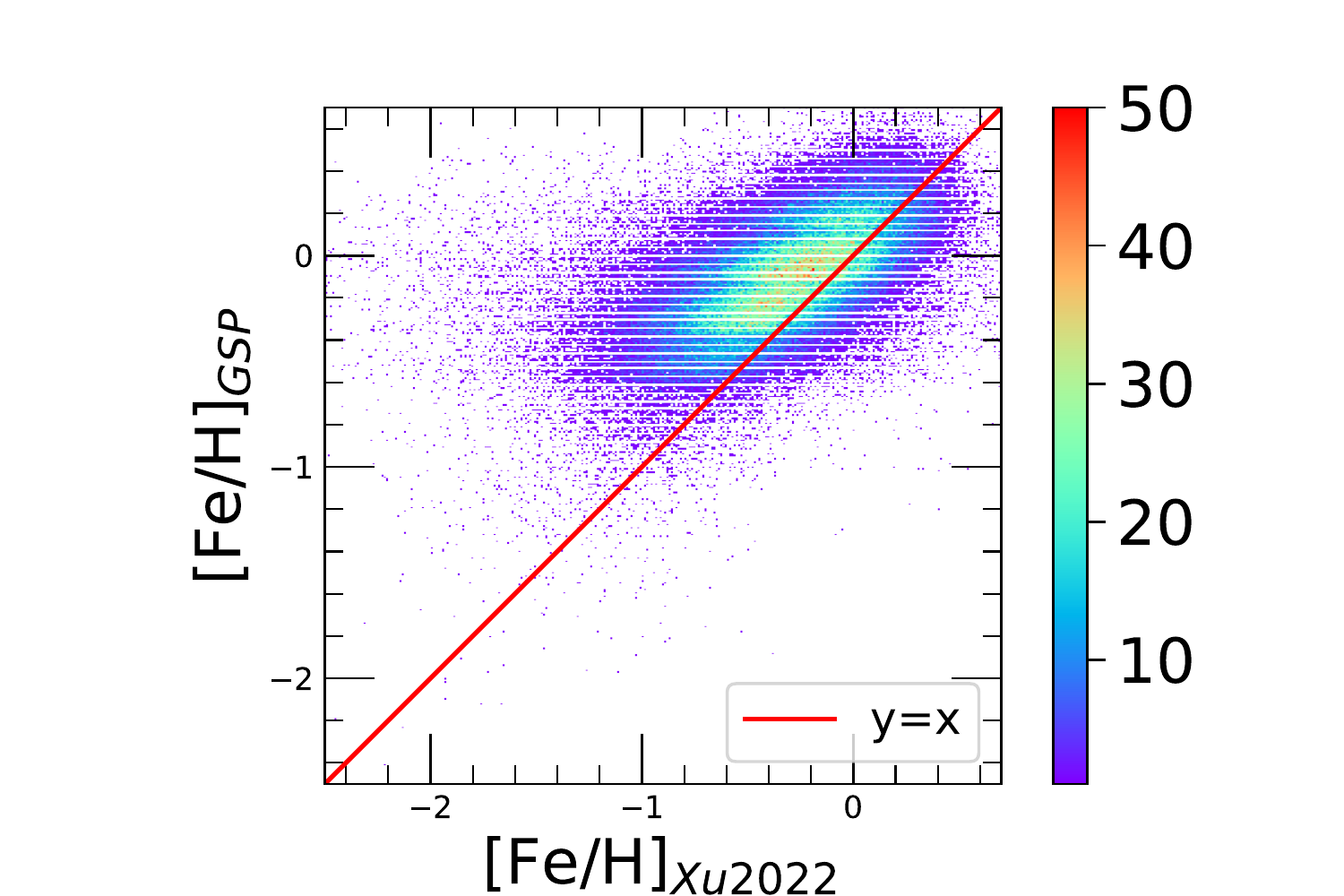}
}
\subfigure{
\includegraphics[width=7cm]{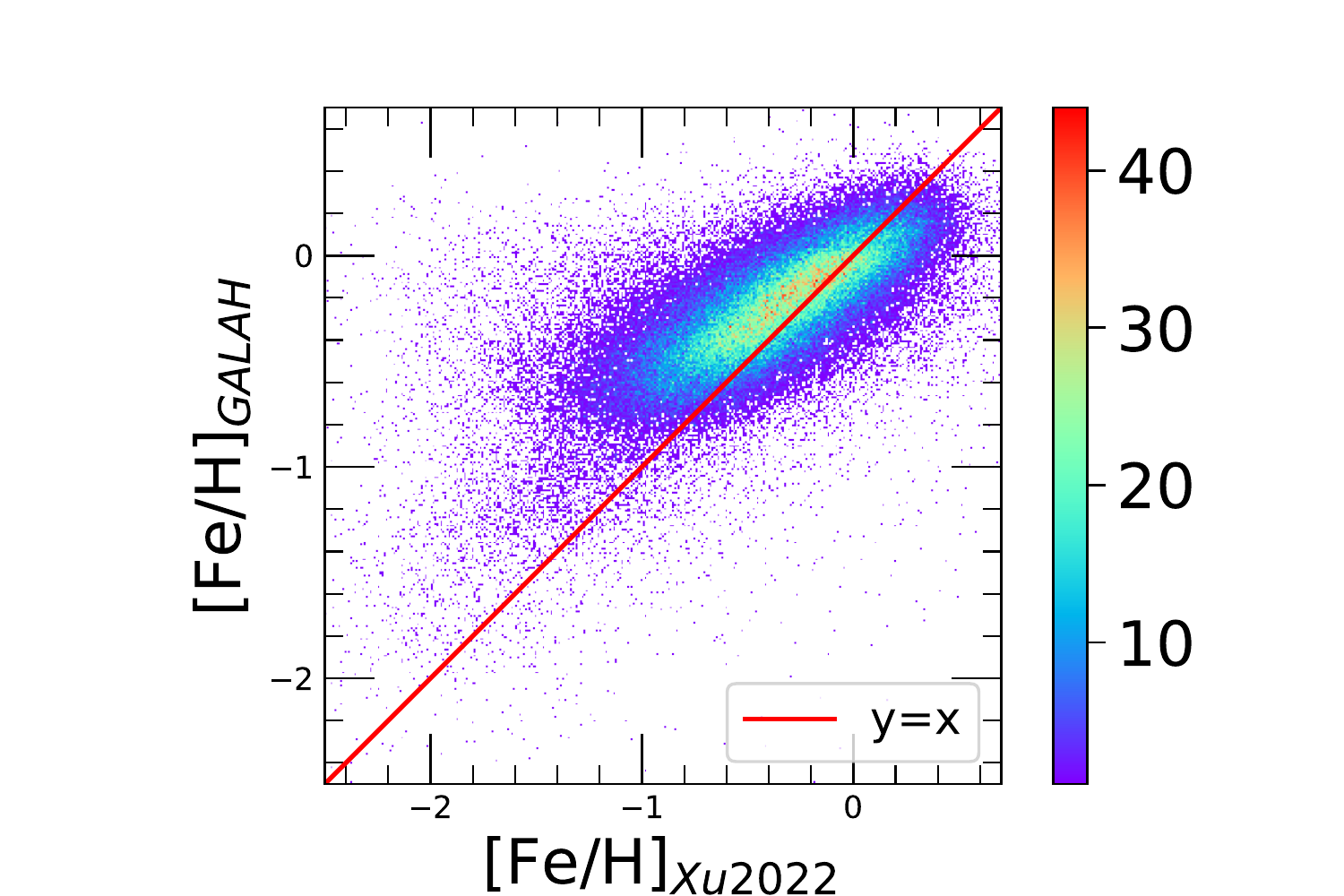}
}

\subfigure{
\includegraphics[width=5cm]{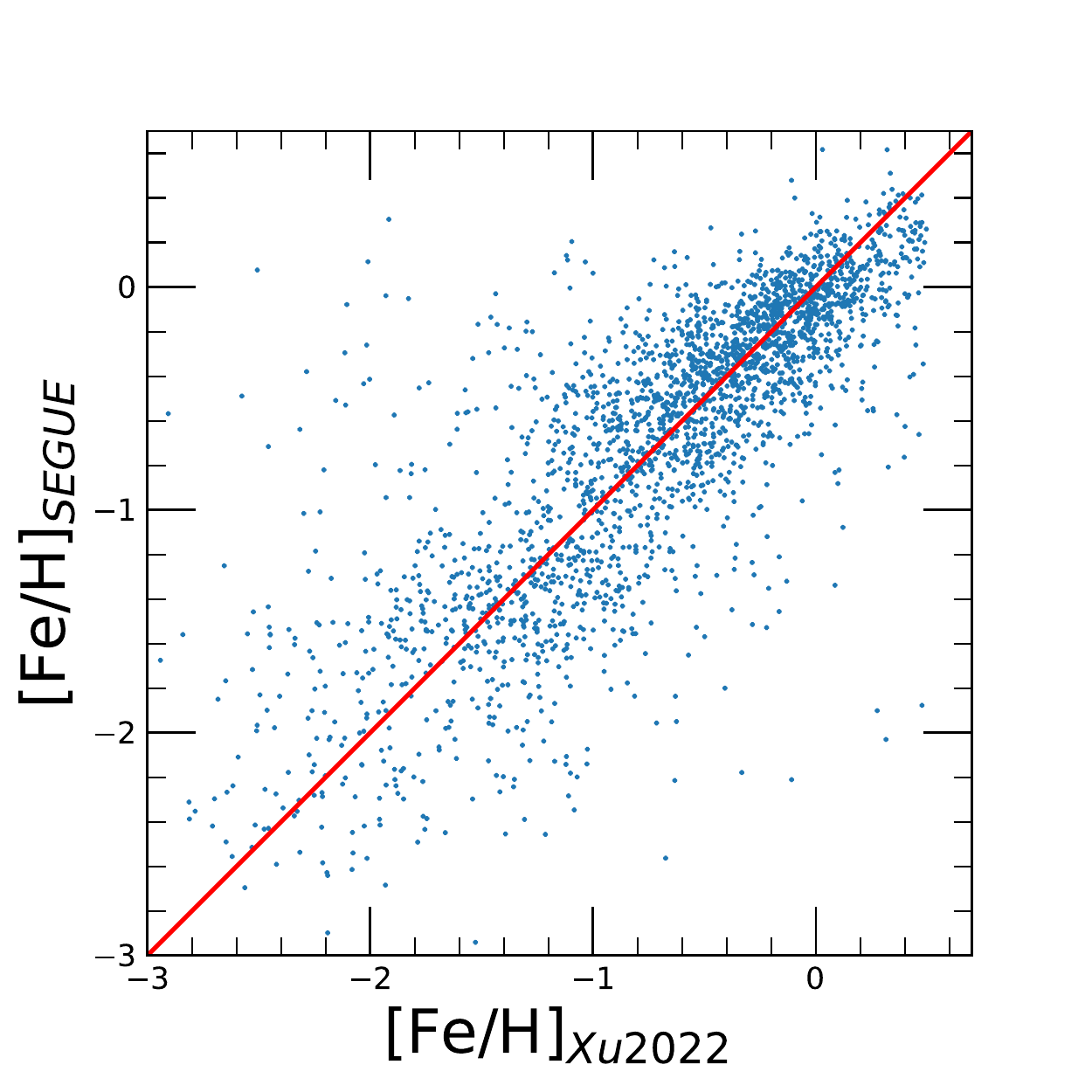}
}
\subfigure{
\includegraphics[width=5cm]{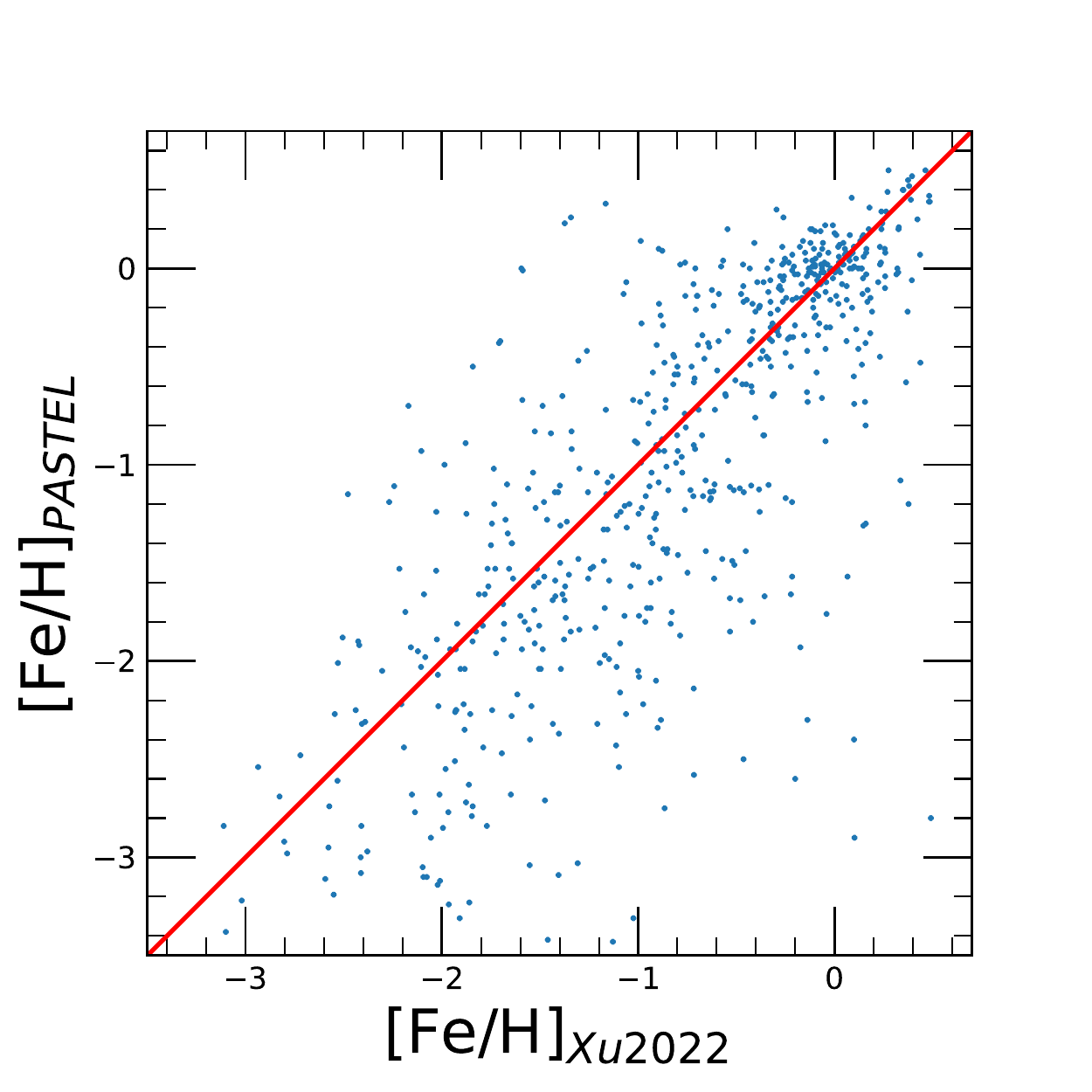}
}
\subfigure{
\includegraphics[width=5cm]{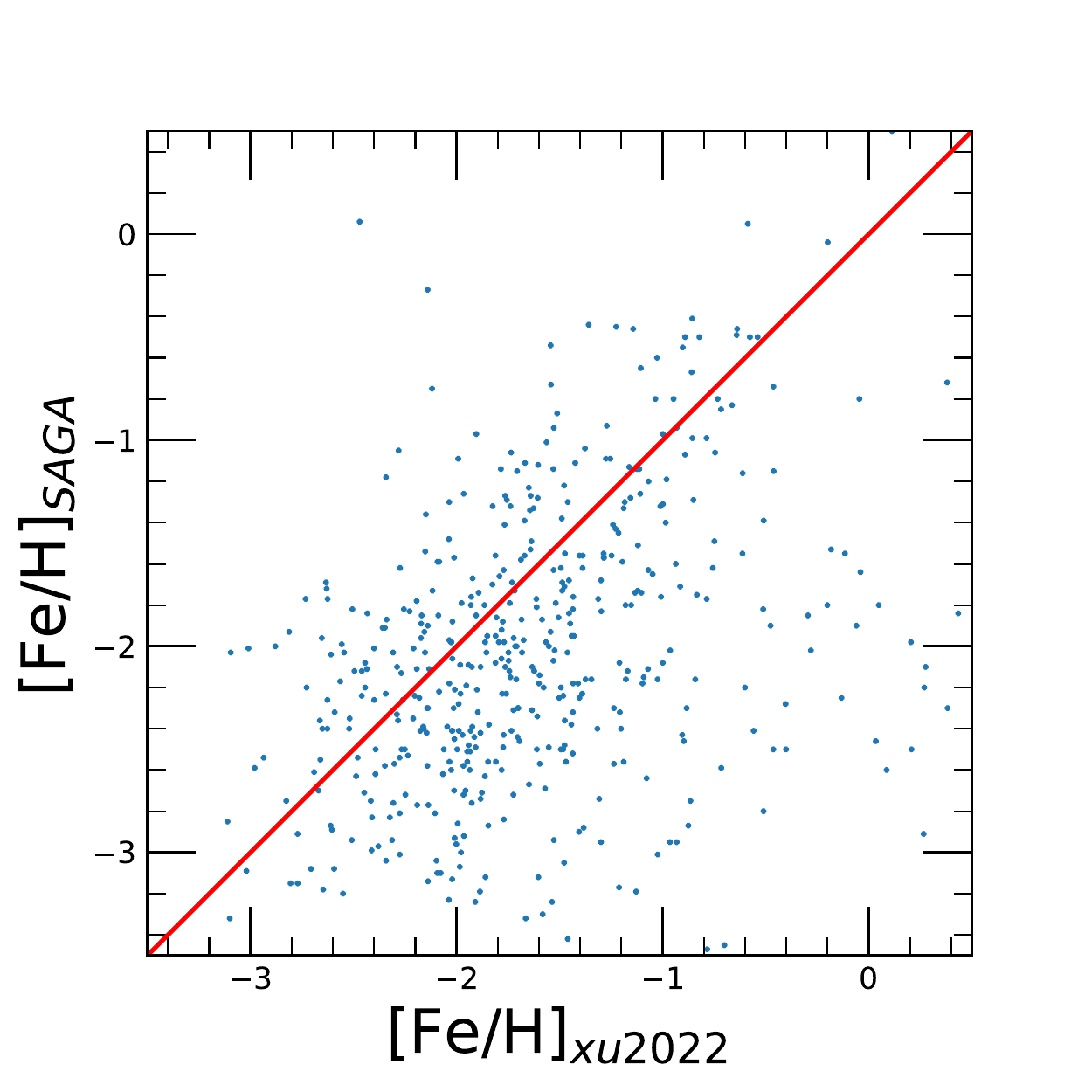}
}
\caption{Comparison of [Fe/H] measurements from  \cite{xu2022} with the Gaia GSP-Spec (top left), GALAH (top right), SEGUE (bottom left), PASTEL (bottom middle), and SAGA (bottom right) catalogs. The red line in each panel is the one-to-one line.  The colors in the top panels indicate number densities.}
\label{fig:compare}
\end{figure*}

We cross-match \cite{xu2022} catalog with five other catalogs: Gaia GSP-Spec, the GALactic Archaeology with HERMES (GALAH) survey (\citealt{martell2016,buder2021}, SEGUE, PASTEL (\citealt{pastel2010,pastel2016}), and the Stellar Abundances for Galactic Archaeology (SAGA; \citealt{saga2008,saga2011,saga2013,saga2017}).
Only giants with $G < 15$ are used in this comparison. The numbers of common sources are 144549, 126338, 2501, 611, and 501 for the Gaia GSP-Spec, GALAH, SEGUE, PASTEL, and SAGA, respectively.
The results are shown in Figure \ref{fig:compare}.
One can see that the [Fe/H] measurements are  consistent. Note that the GSP-Spec metallicities are systematically higher by about 0.25 dex compared to the \cite{xu2022} results.

\bibliographystyle{aasjournal}
\bibliography{metal_poor}

\end{document}